\documentclass[oneside,twocolumn,showpacs,preprintnumbers,amsmath,amssymb,nofootinbib]{revtex4-1}

\usepackage{amsmath}
\usepackage{amsfonts}
\usepackage{amssymb, multirow}
\usepackage{graphicx}
\usepackage{array}
\usepackage[colorlinks=true,linktocpage=true,linkcolor=red,citecolor=blue]{hyperref}

\usepackage{cleveref}
\usepackage[utf8]{inputenc}
\crefname{section}{§}{§§}

\begin{document}

\title{Hypermultiplet gaugings and supersymmetric solutions\\[2mm] from 11D and massive IIA supergravity on H$^{(p,q)}$ spaces\\[2mm]}

\author{Adolfo Guarino}

\affiliation{
\vspace{5mm}
Universit\'e Libre de Bruxelles (ULB) and International Solvay Institutes,\\
Service  de Physique Th\'eorique et Math\'ematique, \\
Campus de la Plaine, CP 231, B-1050, Brussels, Belgium.
}

\begin{abstract}

Supersymmetric $\,\textrm{AdS}_{4}\,$, $\,\textrm{AdS}_{2} \times \Sigma_{2}\,$ and asymptotically AdS$_{4}$ black hole solutions are studied in the context of non-minimal $\,\mathcal{N}=2\,$ supergravity models involving three vector multiplets (\mbox{STU-model}) and Abelian gaugings of the universal hypermultiplet moduli space. Such models correspond to \mbox{consistent} subsectors of the $\,\textrm{SO}(p,q)\,$ and $\,\textrm{ISO}(p,q)\,$ gauged maximal supergravities that arise from the reduction of 11D and massive IIA supergravity on $\,\textrm{H}^{(p,q)}\,$ spaces down to four dimensions. A unified description of all the models is provided in terms of a square-root prepotential and the gauging of a duality-hidden symmetry pair of the universal hypermultiplet. Some aspects of \mbox{M-theory} and massive IIA holography are mentioned in passing.

\end{abstract}

\pacs{04.65.+e, 11.25.Mj \\[1mm] e-mail: aguarino@ulb.ac.be}

\maketitle

\section{Motivation}
\label{Sec:Motivation}

Asymptotically anti-de Sitter (AdS$_{4}$) black holes in minimal $\,\mathcal{N}=2\,$ gauged supergravity have recently been connected to a universal renormalisation group (RG) flow for a large class of three-dimensional $\,\mathcal{N}=2\,$ superconformal field theories (SCFTs) using holography \cite{Azzurli:2017kxo}. The relevant (universal) AdS$_{4}$ black hole is static, extremal (thus $\,T=0\,$) and of Reissner--Nordstr\"om (R--N) type with zero mass and a hyperbolic horizon $\,\Sigma_{2}=\mathbb{H}^2\,$ \cite{Caldarelli:1998hg}. The space-time metric takes the form
\begin{equation}
\label{metric_universal_BH}
d s^2 = - e^{2U} \, d t^2 + e^{-2U} \, d r^2 + r^2 \, d\Omega_{\Sigma_2} \ ,
\end{equation}
with
\begin{equation}
\label{metric_function_universal_BH}
e^{2 U} =  \left( \frac{r}{L_{\textrm{AdS}_{4}}} -  \frac{L_{\textrm{AdS}_{4}}}{2 \, r} \right)^2 \ ,
\end{equation}
and $\,d\Omega_{\Sigma_2}=d \theta^2 + \sinh^2(\theta) \, d \phi^2\,$ being the Riemann \mbox{surface} element on $\,\Sigma_{2}=\mathbb{H}^2\,$. The metric asymptotes an AdS$_{4}$ geometry with radius $\,L_{\textrm{AdS}_{4}}\,$ when $\,{r \rightarrow \infty}\,$, and conforms to $\,{\textrm{AdS}_{2} \times \mathbb{H}^2}\,$ in the near-horizon region $\,r \rightarrow r_{h}=L_{\textrm{AdS}_{4}}/\sqrt{2}\,$ after a shift of the radial coordinate $\,r \rightarrow r \,+\, r_{h}\,$ and the identification $\,L_{\textrm{AdS}_{2}}^2 = \tfrac{1}{4} \, L_{\textrm{AdS}_{4}}^2\,$. 
This black hole is a solution of the equations of motion that follow from the cosmological Einstein--Maxwell \mbox{Lagrangian}
\begin{equation}
\label{Lagrangian_N2_gravity_multiplet}
\mathcal{L} =  \left( \tfrac{1}{2} \, R - V \right) *  1 - \tfrac{1}{2}  \,  \mathcal{H} \wedge * \mathcal{H} \ .
\end{equation}
Endowing the Lagrangian (\ref{Lagrangian_N2_gravity_multiplet}) with $\,\mathcal{N}=2\,$ local supersymmetry requires the cosmological constant to be negative and also a mass term for the gravitini fields in the theory \cite{Freedman:1976aw}. Furthermore, supersymmetry fixes the cosmological constant to $\,{V=-3  \, L_{\textrm{AdS}_{4}}^{-2}=-3  \, |\mu|^2}\,$ in terms of the mass $\,\mu\,$ of a (single) complex gravitino, and renders the black hole magnetically charged \cite{Caldarelli:1998hg} (see also \cite{Romans:1991nq}), \textit{i.e.} $\,{\mathcal{H} = p \, \sinh(\theta) \,   d\theta \wedge d\phi}\,$, with the constant flux $\,p\,$ being also set by supersymmetry. 

The AdS$_{4}$ black hole described above is gauge/gravity dual to a universal RG flow in field theory \cite{Azzurli:2017kxo}. This is an RG flow across dimensions\footnote{See \cite{Bobev:2017uzs} for a generalisation to asymptotically AdS black branes in various dimensions.} from a three-dimensional $\,\mathcal{N}=2\,$ SCFT (dual to the asymptotic AdS$_{4}$ geometry) placed on $\,\mathbb{H}^{2}\,$ and with a topological twist along the exact superconformal \mbox{R-symmetry},
to a one-dimensional superconformal quantum mechanics (dual to the AdS$_{2}$ factor of the black hole near-horizon geometry). Such a universal RG flow admits various holographic embeddings in eleven-dimensional (11D) supergravity \cite{Cremmer:1978km,Gauntlett:2007ma}, the low-energy limit of M-theory, and ten-dimensional massive IIA supergravity \cite{Romans:1985tz}. Specific examples have been studied in the context of ABJM theory \cite{Aharony:2008ug} and GJV/SYM-CS duality \cite{Guarino:2015jca} (and its generalisation of \cite{Fluder:2015eoa}) involving reductions of M-theory and massive IIA strings on various compact spaces \cite{Azzurli:2017kxo}. More concretely, when placing the SCFTs on $\,\textrm{S}^1 \times \Sigma_{2}\,$, a counting of supersymmetric ground states using the topologically twisted index of \cite{Benini:2015noa} at large $\,N\,$ was shown to exactly reproduce the Bekenstein—Hawking entropy associated with the AdS$_4$ black hole in (\ref{metric_universal_BH})-(\ref{metric_function_universal_BH}).

Extensions to non-minimal $\,\mathcal{N}=2\,$ supergravity coupled to matter multiplets, namely, $\,n_{v}\,$ vector multiplets and $\,n_{h}\,$ hypermultiplets, have also been investigated in the context of M-theory \cite{Benini:2015eyy,Benini:2016rke,Cabo-Bizet:2017jsl} and massive IIA strings \cite{Hosseini:2017fjo,Benini:2017oxt}. In the M-theory case, the relevant gauged supergravity is the so-called STU-model with $\,(n_{v},n_{h})=(3,0)\,$ and Fayet-Iliopoulos gaugings with equal gauging parameters $\,g_{\Lambda}=g\,$. This model arises from 11D supergravity when reduced on a compact seven-sphere $\textrm{H}^{(8,0)}\equiv \textrm{S}^7$ \cite{deWit:1986oxb} down to a four-dimensional SO(8) gauged maximal supergravity \cite{deWit:1982ig}, and further truncated to its $\textrm{U}(1)^4$ invariant subsector \cite{Duff:1999gh,Cvetic:1999xp}. The gauging parameter $\,g\,$ is then identified with the inverse radius of $\,\textrm{S}^7$. In the massive IIA case, the relevant gauged supergravity has $\,(n_{v},n_{h})=(3,1)\,$ and involves Abelian gaugings of the universal hypermultiplet moduli space with gauging parameters $\,g\,$ and $\,m\,$. This model arises from ten-dimensional massive IIA supergravity when reduced on a compact six-sphere $\textrm{H}^{(7,0)}\equiv \textrm{S}^6$ \cite{Guarino:2015vca} down to a four-dimensional ISO(7) gauged maximal supergravity \cite{Hull:1984yy,Guarino:2015qaa} of dyonic type \cite{Dall'Agata:2014ita}, and further truncated to its $\textrm{U}(1)^2$ invariant subsector \cite{Guarino:2017eag}. The gauging parameter $\,g\,$ is again identified with the inverse radius of $\,\textrm{S}^6\,$ whereas $\,m\,$ corresponds to the Romans mass parameter. Supersymmetric AdS$_{4}$ black holes generalising the one in (\ref{metric_universal_BH})-(\ref{metric_function_universal_BH}) have been constructed in such M-theory \cite{Cacciatori:2009iz} and massive~IIA \cite{Guarino:2017eag,Guarino:2017pkw} non-minimal $\,\mathcal{N}=2\,$ supergravity models.

In this note we build upon the above results and study non-minimal $\,\mathcal{N}=2\,$ supergravity models that arise from 11D and massive IIA supergravity when reduced on $\,\textrm{H}^{(p,q)}=\textrm{SO}(p,q)/\textrm{SO}(p-1,q)\,$ homogeneous spaces with various $\,(p,q)\,$ signatures. In the case of 11D supergravity, the zero-mass sector recovers an electrically-gauged maximal supergravity in four dimensions with $\,\textrm{SO}(p,q)\,$ gauge group and $\,{p+q=8}\,$ \cite{Hull:1988jw}. In the case of ten-dimensional massive IIA supergravity, the reduction on $\,\textrm{H}^{(p,q)}\,$ spaces down to four dimensions yields a dyonically-gauged maximal supergravity with $\,\textrm{ISO}(p,q)=\textrm{CSO}(p,q,1)\,$ gauge group and $\,p+q=7\,$ \cite{Guarino:2015tja,Cassani:2016ncu,Inverso:2016eet}. We provide a systematic characterisation and a unified description of the $\,\textrm{U}(1)^2\,$ invariant sectors associated with such \mbox{M-theory} and massive IIA reductions\footnote{Einstein-scalar systems associated with $\,{\mathbb{Z}_{2} \times \textrm{SO(3)}}\,$ and $\,\textrm{SU}(3)\,$ invariant sectors of such $\,\textrm{H}^{(p,q)}\,$ reductions were presented in \cite{Guarino:2015tja}.}. The resulting models describe non-minimal $\,\mathcal{N}=2\,$ supergravity with $\,(n_{v},n_{h})=(3,1)\,$ and involve Abelian gaugings of the universal hypermultiplet moduli space. Supersymmetric $\,\textrm{AdS}_{4}\,$, $\,\textrm{AdS}_{2} \times \Sigma_{2}\,$ and universal $\,\textrm{AdS}_{4}\,$ black hole solutions are systematically studied which, upon uplifting to eleven- or ten-dimensional backgrounds,  are of interest for \mbox{M-theory} and massive IIA holography.

\section{$\mathcal{N}=2$ supergravity models}
\label{sec:N=2}

The bosonic sector of $\,\mathcal{N}=2\,$ supergravity coupled to $\,n_{v}\,$ vector multiplets and $\,n_{h}\,$ hypermultiplets is described by a Lagrangian of the form \cite{deWit:2005ub}
\begin{equation}
\label{Lagrangian_N2}
\begin{array}{lll}
\mathcal{L} &=&   \left( \frac{R}{2} - V \right) *  1 - K_{i\bar{j}} \, Dz^{i} \wedge * D\bar{z}^{\bar{j}} - h_{uv} \, Dq^{u} \wedge * Dq^{v}   \\[2mm]
&+& \frac{1}{2} \, \mathcal{I}_{\Lambda \Sigma} \,  \mathcal{H}^\Lambda \wedge * \mathcal{H}^\Sigma +  \frac{1}{2} \,  \mathcal{R}_{\Lambda \Sigma} \, \mathcal{H}^\Lambda \wedge \mathcal{H}^\Sigma \\[2mm]
&+& \frac{1}{2} \Theta^{\Lambda \alpha}\,  \mathcal{B}_{\alpha} \wedge d \tilde{\mathcal{A}}_{\Lambda} + \frac{1}{8} \, \Theta^{\Lambda \alpha} \, \Theta_{\Lambda}{}^{\beta} \, \mathcal{B}_{\alpha} \wedge \mathcal{B}_{\beta} \ .
\end{array}
\end{equation}
In this note we focus on models with three vector multiplets and the universal hypermultiplet \cite{Cecotti:1988qn}, namely, $\,(n_{v}, n_{h})=(3,1)\,$.

The three complex scalars $\,z^{i}\,$ in the vector multiplets $(i=1,2,3)$ serve as coordinates in the special K\"ahler (SK) manifold $\,\mathcal{M}_{\textrm{SK}}=[\textrm{SU}(1,1)/\textrm{U}(1)]^{3}\,$. The metric on this manifold is given by 
\begin{equation}
\label{dSK}
ds_{\textrm{SK}}^2 = K_{i\bar{j}} \, dz^{i} d\bar{z}^{\bar{j}} = \frac{1}{4} \sum_{i}\frac{dz^{i} \, d\bar{z}^{\bar{i}}}{(\textrm{Im} z^{i})^2}  \ ,
\end{equation}
where $\,K_{i \bar{j}} = \partial_{z^{i}} \partial_{\bar{z}^{\bar{j}}} K\,$ and $\,{K=-\log(i \left\langle X, \bar{X} \right\rangle)}\,$ is the real K\"ahler potential. The latter is expressed in terms of an $\,\textrm{Sp}(8)\,$ symplectic product $\,\left\langle X, \bar{X} \right\rangle=X^{M} \Omega_{MN} \bar{X}^{N}=X_{\Lambda}\bar{X}^{\Lambda}-X^{\Lambda}\bar{X}_{\Lambda}\,$ of holomorphic sections $\,X^{M}(z^{i})=(X^{\Lambda},F_{\Lambda})\,$ that satisfy $\,F_{\Lambda} = \partial \mathcal{F}/\partial X^{\Lambda}\,$ ($\Lambda=0,1, 2, 3$) for a homogeneous prepotential of degree-two $\,\mathcal{F}(X^{\Lambda})\,$. Our choice of sections
\begin{equation}
\label{Xsections_nv=3&nh=1}
X^{M} =  (-z^{1} z^{2} z^{3},-z^{1},-z^{2},-z^{3},1, z^{2} z^{3}, z^{3} z^{1}, z^{1} z^{2}) \ ,
\end{equation}
is compatible with a square-root prepotential
\begin{equation}
\label{F_prepot_nv=3&nh=1}
\mathcal{F} = -2 \sqrt{X^0 \, X^1 \, X^2 \, X^3} \ ,
\end{equation}
and restricts the range of the $\,z^{i}\,$ scalars to the K\"ahler cone
\begin{equation}
\label{Kahler_cone}
i \left\langle X, \bar{X} \right\rangle = 8 \, \textrm{Im} z^{1} \, \textrm{Im} z^{2} \, \textrm{Im} z^{3} > 0 \ .
\end{equation}
The condition in (\ref{Kahler_cone}) leaves two different domains: either the three $\,\textrm{Im} z^{i}\,$ are positive, or two of them are negative and the third one is positive.

The kinetic terms and the generalised theta angles for the vector fields in (\ref{Lagrangian_N2}) are given by $\,\mathcal{R}_{\Lambda\Sigma}=\textrm{Re}(\mathcal{N}_{\Lambda\Sigma})\,$ and $\,\mathcal{I}_{\Lambda\Sigma}=\textrm{Im}(\mathcal{N}_{\Lambda\Sigma})\,$ in terms of a complex matrix
\begin{equation}
\label{N_matrix}
\mathcal{N}_{\Lambda \Sigma} = \bar{F}_{\Lambda \Sigma}+ 2 i \frac{\textrm{Im}(F_{\Lambda \Gamma})X^{\Gamma}\,\,\textrm{Im}(F_{\Sigma \Delta})X^{\Delta}}{\textrm{Im}(F_{\Omega \Phi})X^{\Omega}X^{\Phi}} \ ,
\end{equation}
with $\,F_{\Lambda \Sigma}=\partial_{\Lambda}\partial_{\Sigma} \mathcal{F}\,$. They can be used to define a symmetric, real and negative-definite scalar matrix
\begin{equation}
\label{M_matrix}
\mathcal{M}(z^{i}) = \left( 
\begin{array}{cc}
\mathcal{I} + \mathcal{R} \mathcal{I}^{-1} \mathcal{R}   & -\mathcal{R} \mathcal{I}^{-1} \\
- \mathcal{I}^{-1} \mathcal{R} & \mathcal{I}^{-1}
\end{array}
\right) \ ,
\end{equation}
that will appear later on in Sec.~\ref{sec:AdS2}. The vector field strengths are given by
\begin{equation}
\label{H_field_strengths}
\mathcal{H}^{\Lambda} = d \mathcal{A}^{\Lambda} - \tfrac{1}{2} \, \Theta^{\Lambda \alpha} \, \mathcal{B}_{\alpha} \ ,
\end{equation}
and incorporate a set of tensor fields $\,\mathcal{B}_{\alpha}\,$ where $\,\alpha\,$ is a collective index running over the isometries of the scalar manifold that are gauged. Importantly, the tensor fields enter the Lagrangian in (\ref{Lagrangian_N2}) provided that the couplings $\,\Theta^{\Lambda \alpha}  \neq 0\,$, namely, if magnetic charges (see eq.(\ref{covariant_derivatives}) below) are present in the theory \cite{deWit:2005ub}. Tensor fields will play a role in the case of massive IIA reductions on $\,\textrm{H}^{(p,q)}\,$ spaces as magnetic charges are induced by the \mbox{Romans} mass parameter.

The quaternionic K\"ahler (QK) manifold spanned by the four real scalars $\,q^{u}=(\,\phi\,,\sigma\,,\,\zeta\,,\,\tilde{\zeta}\,)\,$ in the universal hypermultiplet is $\,{\mathcal{M}_{\textrm{QK}}=\textrm{SU}(2,1)/\textrm{SU}(2)\times \textrm{U}(1)}\,$. The metric on this QK space reads
\begin{equation}
\label{dsQK_1}
\begin{array}{lll}
ds_{\textrm{QK}}^2 &=&   d \phi \,  d \phi + \frac{1}{4} e^{4 \phi} \left[ d \sigma + \tfrac{1}{2}  \, \vec{\zeta} \,  \mathbb{C} \, d \vec{\zeta}  \right] \, \left[ d \sigma + \tfrac{1}{2}  \, \vec{\zeta} \,  \mathbb{C} \, d \vec{\zeta} \right] \\
&+& \frac{1}{4} \, e^{2\phi} \,   d \vec{\zeta} \,  d\vec{\zeta}   \ ,
\end{array}
\end{equation}
with $\,\mathbb{C}=\left(  \begin{array}{cc} 0 & 1 \\ - 1 & 0\end{array} \right)\,$ and $\,\vec{\zeta}=(\zeta,\tilde{\zeta})\,$. In this note we are exclusively interested in gauging Abelian isometries of $\,{\mathcal{M}_{\textrm{QK}}}\,$ as dictated by an embedding tensor $\,\Theta_{M}{}^{\alpha}\,$ \cite{deWit:2005ub}. They have associated Killing vectors $\,k^{u}_{\alpha}\,$ and yield covariant derivatives in (\ref{Lagrangian_N2}) of the form
\begin{equation}
\label{covariant_derivatives}
Dz^{i} = d z^{i} 
\hspace{5mm} \textrm{ and } \hspace{5mm} 
D q^{u} = d q^{u} - \mathcal{A}^{M} \, \Theta_{M}{}^{\alpha} \,  k_{\alpha}^{u} \ .
\end{equation}

Lastly it turns also convenient to introduce symplectic Killing vectors $\,{\mathcal{K}_{M}  \equiv \Theta_{M}{}^{\alpha} \, k_{\alpha}}\,$ and moment maps $\,{\mathcal{P}_{M}^{x}  \equiv \Theta_{M}{}^{\alpha} \, P_{\alpha}^{x}}\,$ in order to maintain symplectic covariance \cite{Klemm:2016wng}. The scalar potential in (\ref{Lagrangian_N2}) can then be expressed as \cite{Michelson:1996pn,deWit:2005ub}
\begin{equation}
\label{VN2}
\begin{array}{lll}
V &=&  4 \,  \mathcal{V}^{M}  \, \bar{\mathcal{V}}^{N}    \, \mathcal{K}_{M}{}^{u}  \, h_{uv} \,  \mathcal{K}_{N}{}^{v} \\
&+& \mathcal{P}^{x}_{M} \, \mathcal{P}^{x}_{N} \left( K^{i\bar{j}} \, D_{i}\mathcal{V}^{M} \, D_{\bar{j}} \bar{\mathcal{V}}^{N}  - 3 \, \mathcal{V}^{M} \, \bar{\mathcal{V}}^{N} \right) \ ,
\end{array}
\end{equation}
in terms of rescaled sections $\,\mathcal{V}^{M} \equiv e^{K/2} \, X^{M}\,$ and their K\"ahler derivatives $\,D_{i}\mathcal{V}^{M}=\partial_{z^{i}} \mathcal{V}^{M} + \frac{1}{2} (\partial_{z^{i}} K)\mathcal{V}^{M}\,$.

\section{Abelian hypermultiplet gaugings from reductions on $\textrm{H}^{(p,q)}$}
\label{sec:gaugings}

The metric (\ref{dsQK_1}) on the special QK manifold associated with the universal hypermultiplet lies in the image of a \mbox{c-map} \cite{deWit:1990na,deWit:1992wf,deWit:1993rr} with a trivial special K\"ahler base. There are three isometries $\,k_{\alpha}=\{ k_{\sigma} \, , \, \widehat{k}_{\sigma} \, , \, k_{\mathbb{U}} \}\,$ of (\ref{dsQK_1}) that play a role in our reductions of 11D and massive IIA supergravity on $\,\textrm{H}^{(p,q)}\,$ spaces. The isometry $\, k_{\sigma} \,$ corresponds to a model-independent (axion shift) duality symmetry. On the contrary, the isometry $\, \widehat{k}_{\sigma}\,$ corresponds to a hidden symmetry. Together, they form a duality-hidden symmetry pair (conjugate roots of the global symmetry algebra) given by
\begin{equation}
\label{Killing_vectors_sigma}
\begin{array}{lll}
k_{\sigma}&=&-\partial_{\sigma} \ , \\[2mm]
\widehat{k}_{\sigma}&=& \sigma \, \partial_{\phi} - (\sigma^2 - e^{-4 \phi} - U) \, \partial_{\sigma} \\
&& - \left[ \, \sigma \vec{\zeta} - \mathbb{C}  \, (\partial_{\vec{\zeta}} U)  \, \right]^{T} \,  \partial_{\vec{\zeta}} \ ,
\end{array}
\end{equation}
with
\begin{equation}
\label{Ufunc}
U = \frac{1}{16}  \,  |\vec{\zeta}|^4 \, + \, \frac{1}{2}   \, e^{-2 \phi } \, |\vec{\zeta}|^2  \ .
\end{equation}
The remaining isometry $\,k_{\mathbb{U}}\,$ corresponds to a model-dependent duality symmetry and, for the M-theory and massive IIA models studied in this note, it is given by
\begin{equation}
\label{Killing_vectors_U}
\begin{array}{lll}
k_{\mathbb{U}}&=& \tilde{\zeta} \, \partial_{\zeta} - \zeta \, \partial_{\tilde{\zeta}}  \ .
\end{array}
\end{equation}
Note that, while $\,\widehat{k}_{\sigma} + k_{\sigma}\,$ and $\,k_{\mathbb{U}}\,$ are compact Abelian isometries of (\ref{dsQK_1}), the combination $\,\widehat{k}_{\sigma} - k_{\sigma}\,$ turns to be non-compact. 

The triplet of moment maps $\,P_{\alpha}^{x}\,$ associated to the Killing vectors in (\ref{Killing_vectors_sigma}) and (\ref{Killing_vectors_U}) can be obtained from the general construction of \cite{Erbin:2014hsa}. The Killing vectors in (\ref{Killing_vectors_sigma}) have associated moment maps of the form
\begin{equation}
P^{x}_{\sigma} =
\left(
\begin{array}{c}
0 \\[2mm]
0  \\[2mm]
-\frac{1}{2} \, e^{2 \phi } 
\end{array}
\right) \ ,
\end{equation}
and
\begin{equation}
\widehat{P}^{x}_{\sigma} =
\left(
\begin{array}{c}
- e^{-\phi } \, \tilde{\zeta } + e^{\phi }\,  \left( \, - \sigma \, \zeta  +\frac{1}{4}  \,  |\vec{\zeta}|^2 \,\tilde{\zeta } \, \right) \\[2mm]
e^{-\phi } \, \zeta   + e^{\phi }\,  \left( \, - \sigma \, \tilde{\zeta } - \frac{1}{4}   \,  |\vec{\zeta}|^2  \, \zeta \, \right) \\[2mm]
-\frac{1}{2} \, e^{-2 \phi } -\frac{1}{2} \, e^{2 \phi } \, \sigma ^2 -\frac{1}{32} \, e^{2 \phi } \,  |\vec{\zeta}|^4   + \frac{3}{4} \,  |\vec{\zeta}|^2 
\end{array}
\right) \ ,
\end{equation}
whereas the moment maps for the isometry in (\ref{Killing_vectors_U}) take the simpler form
\begin{equation}
P^{x}_{\mathbb{U}} =
\left(
\begin{array}{c}
e^{\phi } \, \tilde{\zeta } \\[2mm]
- e^{\phi } \, \zeta \\[2mm]
1 - \frac{1}{4} \, e^{2 \phi } \, |\vec{\zeta}|^2
\end{array}
\right) \ .
\end{equation}

\begin{table}[t]
\renewcommand{\arraystretch}{1.5}
\begin{tabular}{!{\vrule width 1.5pt}c!{\vrule width 1pt}c!{\vrule width 1pt}ccc!{\vrule width 1pt}cc!{\vrule width 1.5pt}c!{\vrule width 1pt}c!{\vrule width 1.5pt}}
\noalign{\hrule height 1.5pt}
 \,\,\textsc{model} \,\,                     &  $\,\,g_{0}\,\,$ & $\,\,g_{1}\,\,$ & $\,\,g_{2}\,\,$ & $\,\,g_{3}\,\,$ & $\,\,m_{0}\,\,$ & $\,\,m_{i}\,\,$ & $k_{1}$ & $k_{2}$ \\
\noalign{\hrule height 1pt}
 $\textrm{SO}(8)$   &   $g$ & $g$ & $g$ & $g$ & $0$ & $0$ & $\,\,\widehat{k}_{\sigma} + k_{\sigma} \,\,$   & $\,\,k_{\mathbb{U}}\,\,$  \\
 $\textrm{SO}(7,1)$   &   $-g$ & $g$ & $g$ & $g$ & $0$ & $0$ & $\,\,\widehat{k}_{\sigma} -  k_{\sigma} \,\,$   & $\,\,k_{\mathbb{U}}\,\,$  \\
$\textrm{SO}(6,2)_{a}$   &   $-g$ & $g$ & $g$ & $g$ & $0$ & $0$ & $\,\,\widehat{k}_{\sigma} + k_{\sigma} \,\,$   & $\,\,k_{\mathbb{U}}\,\,$ \\
$\textrm{SO}(6,2)_{b}$   &   $g$ & $g$ & $g$ & $-g$ & $0$ & $0$ & $\,\,\widehat{k}_{\sigma} + k_{\sigma} \,\,$   & $\,\,k_{\mathbb{U}}\,\,$ \\
 $\textrm{SO}(5,3)$   &   $-g$ & $g$ & $g$ & $-g$ & $0$ & $0$ & $\,\,\widehat{k}_{\sigma} - k_{\sigma} \,\,$   & $\,\,k_{\mathbb{U}}\,\,$  \\
 $\textrm{SO}(4,4)$   &   $-g$ & $g$ & $g$ & $-g$ & $0$ &  $0$ & $\,\,\widehat{k}_{\sigma} + k_{\sigma} \,\,$   & $\,\,k_{\mathbb{U}}\,\,$  \\
\noalign{\hrule height 1pt}
 $\textrm{ISO}(7)$   &   $g$ & $g$ & $g$ & $g$ & $m$ &  $0$ & $\,\,k_{\sigma} \,\,$   & $\,\,k_{\mathbb{U}}\,\,$  \\
 $\textrm{ISO}(6,1)$   &   $-g$ & $g$ & $g$ & $g$ & $m$ &  $0$ & $\,\,k_{\sigma} \,\,$   & $\,\,k_{\mathbb{U}}\,\,$  \\
 $\textrm{ISO}(5,2)$   &   $g$ & $g$ & $g$ & $-g$ & $m$ &  $0$ & $\,\,k_{\sigma} \,\,$   & $\,\,k_{\mathbb{U}}\,\,$  \\
 $\textrm{ISO}(4,3)$   &   $-g$ & $g$ & $g$ & $-g$ & $m$  &  $0$& $\,\,k_{\sigma} \,\,$   & $\,\,k_{\mathbb{U}}\,\,$  \\
\noalign{\hrule height 1.5pt}
\end{tabular}
\caption{Embedding tensor (\ref{Theta_tensor_N2}) and gauged isometries for the $\,\mathcal{N}=2\,$ supergravity models. For the sake of definiteness, and without loss of generality, we are taking $\,p \geq q\,$ as well as $\,g>0\,$ and $\,m>0\,$.}
\label{Table:embedding_tensor}
\end{table}

In order to fully determine the $\,\mathcal{N}=2\,$ supergravity model, one must still specify the gauge connection entering the covariant derivatives in (\ref{covariant_derivatives}). This is done by an embedding tensor $\,\Theta_{M}{}^{\alpha}\,$ of the form
\begin{equation}
\label{Theta_tensor_N2}
\Theta_{M}{}^{\alpha} = \left(
\begin{array}{c}
\Theta_{\Lambda}{}^{\alpha} \\[2mm]
\hline\\[-2mm]
\Theta^{\Lambda \, \alpha} 
\end{array}\right)
= \left(
\begin{array}{cc}
g_{0} & 0 \\[2mm]
0 & g_{1} \\[2mm]
0 & g_{2} \\[2mm]
0 & g_{3} \\[2mm]
\hline\\[-2mm]
m_{0} & 0 \\[2mm]
0 & m_{1} \\[2mm]
0 & m_{2} \\[2mm]
0 & m_{3}
\end{array}\right) \ ,
\end{equation}
where the various electric $\,g_{\Lambda}\,$ and magnetic $\ m_{\Lambda}\,$ charges are displayed in Table~\ref{Table:embedding_tensor}. The covariant derivatives in (\ref{covariant_derivatives}) reduce to
\begin{equation}
\label{Dq}
\begin{array}{lll}
D z^{i} &=& d z^{i} \ , \\[2mm]
D q^{u} &=& d q^{u} - (g_{0} \, \mathcal{A}^{0} + m_{0} \, \mathcal{\tilde{A}}_{0}) \, k_{1}^{u} - \mathcal{A}  \, k_{2}^{u} \ ,
\end{array}
\end{equation}
with $\,\mathcal{A}=\sum_{i} g_{i} \, \mathcal{A}^{i} + \sum_{i} m_{i} \, \tilde{\mathcal{A}}_{i}\,$. However, the specific isometries $\,k_{1}\,$ and $\,k_{2}\,$ to be gauged in (\ref{Dq}) depend on the M-theory or massive IIA origin of the models:
\begin{equation}
\label{k1k2isometries}
\begin{array}{lllll}
\textrm{M-theory} : & k_{1} =\widehat{k}_{\sigma} + (-1)^{p q} \,  k_{\sigma}   & \,,\,  & k_{2}=k_{\mathbb{U}}  & , \\[2mm]
\textrm{Massive IIA} : & k_{1} = k_{\sigma} & \, , \, & k_{2}= k_{\mathbb{U}} & . \\[2mm]
\end{array}
\end{equation}
The $\,(-1)^{p q}\,$ sign in $\,k_{1}\,$ for the M-theory models depends on the signature of the $\,\textrm{H}^{(p,q)}\,$ space employed in the \mbox{reduction}. Moreover, as a consequence of (\ref{k1k2isometries}), only duality symmetries $\,k_{\sigma}\,$ and $\,k_{\mathbb{U}}\,$ appear upon reductions of massive~IIA supergravity whereas also the hidden symmetry $\,\widehat{k}_{\sigma}\,$ does it in the reductions of M-theory. Note also that, in the massive IIA case, the resulting gaugings are of dyonic type: they involve both electric and magnetic charges.

\section{Supersymmetric solutions}
\label{Sec:SUSY-solutions}

The M-theory and massive IIA non-minimal $\,{\mathcal{N}=2}\,$ supergravity models presented in the previous section possess various types of supersymmetric solutions.

\subsection{$\textrm{AdS}_{4}$ solutions}

An $\,\textrm{AdS}_{4}\,$ vacuum solution with radius $\,L_{\textrm{AdS}_{4}}\,$ is describing a space-time geometry of the form
\begin{equation}
\label{metric_AdS4}
d s^2 = - \frac{r^2}{L^2_{\textrm{AdS}_{4}}} \, d t^2 + \frac{L^2_{\textrm{AdS}_{4}}}{r^2} \, d r^2 + r^{2} \, d\Omega_{\Sigma_2} \ ,
\end{equation}
where $\,d\Omega_{\Sigma_2}=d \theta^2 + \left( \frac{ \sin \sqrt{\kappa} \, \theta }{ \sqrt{ \kappa } } \right)^2 \, d \phi^2\,$ is the surface element of $\,\Sigma_{2}=\left\lbrace \mathbb{S}^2 \, (\kappa=+1) \, , \,  \mathbb{H}^2 \, (\kappa=-1) \right\rbrace\,$. Being a maximally symmetric solution, only scalars are allowed to acquire a non-trivial and constant vacuum expectation value that extremises the potential in (\ref{VN2}). In addition, preserving $\,{\mathcal{N}=2}\,$ supersymmetry requires the vanishing of all fermionic supersymmetry variations. This translates into the conditions \cite{Louis:2012ux}: 
\begin{equation}
\label{N=2_AdS4_cond}
\begin{array}{rlll}
X^{M}\, \mathcal{K}_{M} &=& 0 & , \\[2mm]
\left[ \,  \partial_{z^{i}} X^{M} + (\partial_{z^{i}} K) X^{M} \, \right] \, \mathcal{P}^{x}_{M} & = & 0 & , \\[2mm]
S_{\mathcal{AB}} \, \epsilon^{\mathcal{B}} &=& \frac{1}{2} \, \mu \, \epsilon^{\ast}_{\mathcal{A}} & ,
\end{array}
\end{equation}
where $\,S_{\mathcal{AB}}=\frac{1}{2} e^{K/2} X^{M}  \mathcal{P}^{x}_{M} (\sigma^{x})_{\mathcal{AB}}\,$ is the gravitino mass matrix,  $\,(\sigma^{x})_{\mathcal{AB}}\,$ are the Pauli matrices and $\,|\mu|=L^{-1}_{\textrm{AdS}_{4}}\,$.

\subsubsection{M-theory}

In the $\,\textrm{SO}(p,q)\,$ models with $\,k_{1}=\widehat{k}_{\sigma} + (-1)^{pq} \,  k_{\sigma}\,$ the first condition in (\ref{N=2_AdS4_cond}) yields 
\begin{equation}
\label{AdS4_SOpq_1}
\sigma = 0 \hspace{4mm} \textrm{ and } \hspace{4mm} \left(e^{-2 \phi} + \frac{1}{4} \, |\vec{\zeta}|^2\right)^2 = (-1)^{pq} \ .
\end{equation}
The last equation in (\ref{AdS4_SOpq_1}) excludes models with $\,{p \, q \in \textrm{odd}}$, namely, the SO(7,1) and SO(5,3) models. For those models with $\,{p \, q \in \textrm{even}}\,$, the first condition in (\ref{N=2_AdS4_cond}) additionally gives  
\begin{equation}
\label{AdS4_SOpq_2}
\left( \sum_{i} g_{i} \, z^{i} + g_{0} \, \prod_{i} z^{i} \right) \, \vec{\zeta} = 0 \ .
\end{equation}
Extremising the scalar potential in (\ref{VN2}) subject to the conditions in (\ref{N=2_AdS4_cond}) imposed by supersymmetry yields two AdS$_{4}$ solutions that preserve a different residual (\mbox{unbroken}) gauge symmetry.

The first solution fixes the scalars $\,z^{i}\,$ in the vector multiplets as
\begin{equation}
\label{AdS4_SOpq_N8_vector}
z^{i}= \pm \,  i \, \dfrac{(g_{0} \, g_{1} \, g_{2} \, g_{3})^{\frac{1}{2}}}{g_{0} \, g_{i}} \ ,
\end{equation}
and involves a trivial configuration of the scalars in the universal hypermultiplet
\begin{equation}
\label{AdS4_SOpq_N8_hyper}
e^{\phi}= 1  \hspace{5mm} \textrm{ and } \hspace{5mm}  |\vec{\zeta}|^2=0 \ .
\end{equation}
This solution possesses a realisation in two models:
\begin{eqnarray}
\label{AdS4_SO8_N8}
\textrm{SO(8)}  & \,\,:\,\, & z^{1}=z^{2}=z^{3}=  i  \ , \\
\label{AdS4_SO44_N8}
\textrm{SO(4,4)}  & \,\,:\,\, & z^{1}=z^{2}=-z^{3}=  -i  \ .
\end{eqnarray}
In both cases
\begin{equation}
\label{LAdS4_SO8SO44_N8}
L^2_{\textrm{AdS}_{4}} = \frac{1}{2 \, g^2} \ ,
\end{equation}
and the two vectors $\,g_{0} \, \mathcal{A}^{0}\,$ and $\,\mathcal{A}\,$ remain massless ($k_{1}=k_{2}=0$), thus preserving the full $\,\textrm{U}(1)_{1} \times \textrm{U}(1)_{2}\,$ gauge symmetry of the models. In the compact $\,\textrm{SO}(8)\,$ case, this solution uplifts to the maximally supersymmetric Freund-Rubin vacuum of 11D supergravity \cite{Freund:1980xh}, and the dual SCFT is identified with ABJM theory \cite{Aharony:2008ug} at low ($\,k=1,2\,$) Chern--Simons levels $\,k\,$ and $\,-k\,$.

The second solution fixes the scalars $\,z^{i}\,$ at the values
\begin{equation}
\label{AdS4_SOpq_N2_vector}
z^{i}= \pm \,  i \, \sqrt{3}  \, \dfrac{(g_{0} \, g_{1} \, g_{2} \, g_{3})^{\frac{1}{2}}}{g_{0} \, g_{i}}   \ ,
\end{equation}
and involves a non-trivial configuration of the universal hypermultiplet independent of the gauging parameters
\begin{equation}
\label{AdS4_SOpq_N2_hyper}
e^{\phi}= \dfrac{\sqrt{3}}{\sqrt{2}}  \hspace{5mm} \textrm{ and } \hspace{5mm}   |\vec{\zeta}|^2=\frac{4}{3} \ .
\end{equation}
This solution has a realisation in the same models as before:
\begin{eqnarray}
\label{AdS4_SO8_N2}
\textrm{SO(8)}  & \,\,:\,\, & z^{1}=z^{2}=z^{3}=  i \, \sqrt{3}   \ , \\
\label{AdS4_SO44_N2}
\textrm{SO(4,4)}  & \,\,:\,\, & z^{1}=z^{2}=-z^{3}=  -i \, \sqrt{3} \  .
\end{eqnarray}
In both cases
\begin{equation}
\label{LAdS4_SO8SO44_N2}
L^2_{\textrm{AdS}_{4}} = \frac{2}{3 \sqrt{3} \, g^2} \ ,
\end{equation}
and only the linear combination of vectors $\,- \, 3 \, g_0 \,  \mathcal{A}^{0}  + \mathcal{A}\,$ remains massless ($k_{1}=k_{2}$) in the $\,\textrm{SO}(8)\,$ and $\,\textrm{SO}(4,4)\,$ models. The associated $\,\textrm{U}(1)\,$ gauge symmetry is thus preserved and to be identified with the \mbox{R-symmetry} of the dual field theory. In the compact $\,\textrm{SO}(8)\,$ case, this solution was studied in \cite{Warner:1983vz,Nicolai:1985hs} and uplifted to a background of 11D supergravity in \cite{Corrado:2001nv,Ahn:2002eh}. Its dual SCFT was identified in \cite{Benna:2008zy} as the infrared fixed point of an RG flow from ABJM theory triggered by an SU(3) invariant mass term in the superpotential.

The fixing of the scalars $\,z^{i}\,$ in the vector multiplets to the values in (\ref{AdS4_SO8_N8})-(\ref{AdS4_SO44_N8}) and (\ref{AdS4_SO8_N2})-(\ref{AdS4_SO44_N2}) is compatible with the K\"ahler cone condition (\ref{Kahler_cone}) for a physically acceptable solution in $\,\mathcal{N}=2\,$ supergravity in four dimensions.

\subsubsection{Massive IIA}

In the $\,\textrm{ISO}(p,q)\,$ models the first condition in (\ref{N=2_AdS4_cond}) yields
\begin{equation}
\label{AdS4_ISOpq}
\left( \sum_{i} g_{i} \, z^{i} \right)  \, \vec{\zeta} = 0
\hspace{4mm} \textrm{ and } \hspace{4mm}
\prod_{i} z^{i} = \frac{m_{0}}{g_{0}} \ .
\end{equation}
The extremisation of the scalar potential in (\ref{VN2}) subject to the supersymmetry conditions in (\ref{N=2_AdS4_cond}) yields this time a unique AdS$_{4}$ solution. It has
\begin{equation}
\label{AdS4_ISOpq_N2_1}
\begin{array}{lll}
(\textrm{Re}z^{i})^3 &=& -\dfrac{1}{8} \, \dfrac{g_1 \, g_2 \, g_3}{g_{i}^3} \, \dfrac{m_{0}}{g_{0}} \ ,\\[4mm]
(\textrm{Im}z^{i})^3 &=& \pm \dfrac{3 \sqrt{3}}{8} \, \dfrac{g_1 \, g_2 \, g_3}{g_{i}^3} \, \dfrac{m_{0}}{g_{0}} \ ,
\end{array}
\end{equation}
for the $\,z^{i}\,$ scalars in the vector multiplets, together with $\, |\vec{\zeta}|^2=0 \,$ and a non-trivial dilaton
\begin{equation}
\label{AdS4_ISOpq_N2_2}
e^{6 \phi}= 8 \,\,  \dfrac{g_1 \, g_2 \, g_3}{g_0 \, m_{0}^2} \ ,
\end{equation}
in the universal hypermultiplet. Note that (\ref{AdS4_ISOpq_N2_2}) requires $\,g_0 \, g_1 \, g_2 \, g_3 > 0\,$ which is satisfied only by the ISO(7) and ISO(4,3) models (see Table~\ref{Table:embedding_tensor}). The corresponding solutions are given by
\begin{eqnarray}
\label{AdS4_ISO7_N2}
\textrm{ISO(7)}  & \,\,:\,\, & z^{1}=z^{2}=z^{3}= \left(\tfrac{m}{g}\right)^{\frac{1}{3}} \, e^{i \frac{2\pi}{3}}  \ , \\
\label{AdS4_ISO43_N2}
\textrm{ISO(4,3)}  & \,\,:\,\, & z^{1}=z^{2}=-z^{3}= \left(\tfrac{m}{g}\right)^{\frac{1}{3}}  e^{-i \frac{2\pi}{3}} \ ,
\end{eqnarray}
together with
\begin{equation}
\label{phi_AdS4_ISO7_ISO43_N2}
e^{\phi}= \sqrt{2} \, \left(\frac{g}{m}\right)^{\frac{1}{3}} \ .
\end{equation}
Both configurations (\ref{AdS4_ISO7_N2})-(\ref{AdS4_ISO43_N2}) satisfy the K\"ahler cone condition (\ref{Kahler_cone}) and yield 
\begin{equation}
\label{LAdS_4_ISO7_ISO43}
L^2_{\textrm{AdS}_{4}} = \tfrac{1}{\sqrt{3}} \, g^{-\frac{7}{3}}\, m^{\frac{1}{3}} \ .
\end{equation}
The vector $\,\mathcal{A}\,$ remains massless ($k_{2}=0$) and so the $\,\textrm{U}(1)_{2}\,$ gauge symmetry is preserved. This symmetry is holographically identified with the R-symmetry of the dual field theory. In the ISO(7) case, this solution was presented in \cite{Guarino:2015jca,Guarino:2015qaa}\footnote{Note that $\,z^{i}_{\textrm{here}} = e^{i \frac{\pi}{3}} \, z_{\textrm{\cite{Guarino:2015jca}}}^{i}\,$ as a consequence of the sign choice $\,m_{\textrm{here}}=-m_{\textrm{\cite{Guarino:2015jca}}}\,$.} and uplifted to a background of massive IIA supergravity in \cite{Guarino:2015jca,Guarino:2015vca}. The dual three-dimensional SCFT was also identified in \cite{Guarino:2015jca} as a super Chern--Simons-matter theory with simple gauge group $\textrm{SU}(N)$ and level $\,k\,$ given by the Romans mass \mbox{parameter}.

\subsection{$\textrm{AdS}_{2} \times \Sigma_{2}$ solutions}
\label{sec:AdS2}

Let us focus on $\,\textrm{AdS}_{2} \times \Sigma_{2}\,$ vacuum solutions with radii $\,L_{\textrm{AdS}_{2}}\,$ and $\,L_{\Sigma_{2}}\,$. The corresponding space-time geometry is specified by a metric of the form
\begin{equation}
\label{metric_AdS2xSigma2}
d s^2 = - \frac{r^2}{L^2_{\textrm{AdS}_{2}}} \, d t^2 + \frac{L^2_{\textrm{AdS}_{2}}}{r^2} \, d r^2 + L_{\Sigma_{2}}^{2} \, d\Omega_{\Sigma_2} \ .
\end{equation}
The metric in (\ref{metric_AdS2xSigma2}) also describes the horizon of a static and extremal black hole, like the one in (\ref{metric_universal_BH})-(\ref{metric_function_universal_BH}), so these solutions are sometimes referred to as black hole horizon solutions. 

An ansatz for the fields in the vector-tensor sector of the Lagrangian (\ref{Lagrangian_N2}) that is compatible with the space-time symmetries takes the form \cite{Guarino:2017eag} (see \cite{Klemm:2016wng} for a tensor gauge-equivalent choice):
\begin{equation}
\label{Ansatz_vector-tensor}
\begin{array}{llll}
\mathcal{A}^\Lambda &=& \mathcal{A}_t{}^\Lambda(r) \, d t - p^\Lambda \, \frac{ \cos \sqrt{\kappa} \, \theta }{ \kappa } \, d \phi & , \\[2mm]
\tilde{\mathcal{A}}_\Lambda &=& \tilde{\mathcal{A}}_t{}_\Lambda(r) \, d t - e_\Lambda \, \frac{ \cos \sqrt{\kappa} \, \theta }{ \kappa } \, d \phi & , \\[2mm]
\mathcal{B}_{\alpha} &=& b_{\alpha}(r) \, \frac{ \sin \sqrt{\kappa} \, \theta }{ \sqrt{\kappa} } \, d \theta \wedge d \phi & .
\end{array}
\end{equation}
As a consequence of (\ref{H_field_strengths}) and (\ref{Ansatz_vector-tensor}), this sector of the theory is encoded in a vector of charges $\,\mathcal{Q}\,$ of the form $\,\mathcal{Q}^{M} = \left( \mathfrak{p}^{\Lambda}, \mathfrak{e}_{\Lambda} \right)^{T}\,$ with
\begin{equation}
\label{charges_p_e}
\mathfrak{p}^{\Lambda}=p^{\Lambda} - \frac{1}{2} \, \Theta^{\Lambda \, \alpha} \, b_{\alpha}
\hspace{4mm} \textrm{ and } \hspace{4mm}
\mathfrak{e}_{\Lambda}=e_{\Lambda} + \frac{1}{2} \, \Theta_{\Lambda}{}^{\alpha} \, b_{\alpha} \ ,
\end{equation}
which generically depends on vector charges $\,(p^{\Lambda},e_{\Lambda})\,$ as well as on the $\,\theta$-$\varphi\,$ components $\,b_{\alpha}\,$ of the tensor fields in (\ref{Ansatz_vector-tensor}). The latter play a role only in the massive IIA models where the gaugings are of dyonic type.

In the presence of hypermultiplets, quarter-BPS black hole horizon solutions with constant scalars require the set of algebraic equations \cite{Klemm:2016wng} (see also \cite{Halmagyi:2013sla})
\begin{equation}
\label{AdS2xSigma2_system}
\begin{split}
\mathcal{Q} & =   \kappa \, L_{\Sigma_{2}}^{2} \, \Omega \, \mathcal{M} \, \mathcal{Q} ^{x} \, \mathcal{P}^{x} - 4 \, \textrm{Im}(\bar{\mathcal{Z}} \, \mathcal{V}) \ , \\
\dfrac{L_{\Sigma_{2}}^{2}}{L_{\textrm{AdS}_{2}}} & =  -2 \, \mathcal{Z} \, e^{-i \beta} \ , \\[2mm]
\left\langle  \mathcal{K}^{u} , \mathcal{V} \right\rangle & =  0 \ ,
\end{split}
\end{equation}
defined in terms of a central charge $\,\mathcal{Z}(z^{i})=\left\langle \mathcal{Q} , \mathcal{V} \right\rangle\,$, the scalar matrix $\,\mathcal{M}(z^{i})\,$ in (\ref{M_matrix}) and $\,\mathcal{Q}^{x} = \left\langle  \mathcal{P}^{x} , \mathcal{Q} \right\rangle\,$. The phase $\,\beta\,$ is associated with the complex function
\begin{equation}
\label{W_func}
W=e^{U} (\mathcal{Z} + i \, \kappa \, L_{\Sigma_{2}}^2 \, \mathcal{L})= |W| \, e^{i \beta} \ ,
\end{equation}
which depends on the central charge $\,\mathcal{Z}(z^i)\,$ and a superpotential $\,\mathcal{L}(z^{i},q^{u})=\left\langle \mathcal{Q}^{x} \mathcal{P}^{x} , \mathcal{V} \right\rangle\,$. 

With the aim of constructing later on asymptotically AdS$_{4}$ black holes of the universal type in (\ref{metric_universal_BH})-(\ref{metric_function_universal_BH}), we are concentrating on the configurations of the $\,z^{i}\,$ scalars found in the previous section to be compatible with AdS$_{4}$ solutions preserving $\,\mathcal{N}=2\,$ supersymmetry. 

\subsubsection{M-theory}

As for the case of AdS$_{4}$ solutions, the third condition in (\ref{AdS2xSigma2_system}) automatically discards $\,\textrm{AdS}_{2} \times \Sigma_{2}\,$ solutions in the $\,\textrm{SO}(7,1)\,$ and $\,\textrm{SO}(5,2)\,$ models. Nevertheless, black hole horizon solutions with non-zero magnetic charges $\,p^{\Lambda}\,$ exist in all the remaining $\,\textrm{SO}(p,q)\,$ models.

In the case of a trivial configuration of the universal hypermultiplet, various solutions are found for the $\,\textrm{SO}(8)\,$, $\,\textrm{SO}(4,4)\,$ and $\,\textrm{SO}(6,2)_{a,b}\,$ models. Of special interest will be those of the $\,\textrm{SO}(8)\,$ and $\,\textrm{SO}(4,4)\,$ models with the scalars $\,z^{i}\,$ being fixed at the values in (\ref{AdS4_SO8_N8}) and (\ref{AdS4_SO44_N8}), respectively. These solutions are supported by charges of the form
\begin{eqnarray}
\textrm{SO(8)} &  \,\,:\,\, &   p^{0}=-p^{1}=-p^{2}=-p^{3}=\tfrac{1}{4 g} \ , \\
\textrm{SO(4,4)} &  \,\,:\,\, & p^{0}=p^{1}=p^{2}=-p^{3} = - \tfrac{1}{4 g} \ ,
\end{eqnarray}
and both have a hyperbolic horizon $\,\Sigma_{2}=\mathbb{H}^2\,$ ($\kappa = - 1$), a phase $\,\beta=0\,$ in (\ref{W_func}) and radii given by
\begin{equation}
\label{LAdS_2_SO8_SO44_N8}
L^2_{\textrm{AdS}_{2}} = \tfrac{1}{8} \, g^{-2}  \hspace{5mm} , \hspace{5mm}  L^2_{\mathbb{H}^{2}} = \tfrac{1}{4} \, g^{-2} \ .
\end{equation}
For the SO(6,2)$_{a}$ model there are solutions with the scalars $\,z^{i}\,$ being also fixed as in (\ref{AdS4_SO8_N8}) and (\ref{AdS4_SO44_N8}). They are respectively supported by charges of the form 
\begin{eqnarray}
\textrm{SO(6,2)}_{a} &  \,\,:\,\, & p^{0}=\tfrac{1}{g} \,\,\,\,\, , \,\,\,\,\, p^{1}=p^{2}=p^{3}=0 \ ,\\
\textrm{SO(6,2)}_{a} &  \,\,:\,\, & p^{0}=p^{1}=p^{2}=0 \,\,\,\,\, , \,\,\,\,\, p^{3}=\tfrac{1}{g} \ ,
\end{eqnarray}
and both have a spherical horizon $\,\Sigma_{2}=\mathbb{S}^2\,$ ($\kappa = 1$), a phase $\,\beta=0\,$ in (\ref{W_func}) and radii given by
\begin{equation}
L^2_{\textrm{AdS}_{2}} = \tfrac{1}{2} \, g^{-2}
 \hspace{5mm} , \hspace{5mm} 
L^2_{\mathbb{S}^2} = \tfrac{1}{2} \, g^{-2} \ .
\end{equation}
Analogue solutions can also be found in the SO(6,2)$_{b}$ model.

In the case of requiring a non-trivial configuration of the universal hypermultiplet, only the $\,\textrm{SO}(8)\,$ and $\,\textrm{SO}(4,4)\,$ models turn out to accommodate black hole horizon solutions. These have the scalars $\,z^{i}\,$ fixed as in (\ref{AdS4_SO8_N2}) and (\ref{AdS4_SO44_N2}) and the universal hypermultiplet set as in (\ref{AdS4_SOpq_N2_hyper}). The solutions are supported by charges of the form
\begin{eqnarray}
\textrm{SO(8)} &  \,\,:\,\, & p^{0} = -3p^{1} = -3p^{2} = -3p^{3} = \tfrac{1}{2 g} \ , \\
\textrm{SO(4,4)} &  \,\,:\,\, & p^{0} = 3p^{1} = 3p^{2} = -3p^{3} = - \tfrac{1}{2 g} \  ,
\end{eqnarray}
and both have a hyperbolic horizon $\,{\Sigma_{2}=\mathbb{H}^2}\,$ (${\kappa=-1}$), a phase $\,\beta=0\,$ in (\ref{W_func}) and radii given by
\begin{equation}
\label{LAdS_2_SO8_SO44_N2}
 L^2_{\textrm{AdS}_{2}} = \tfrac{1}{6 \sqrt{3}} \, g^{-2}  \hspace{5mm} , \hspace{5mm}    L^2_{\mathbb{H}^2} = \tfrac{1}{3 \sqrt{3}} \, g^{-2} \ .
\end{equation}

Lastly, additional $\,\textrm{AdS}_{2} \times \Sigma_{2}\,$ solutions equivalent to the ones presented above are obtained upon replacing $\,\mathcal{Q} \rightarrow -\mathcal{Q}\,$ and $\,\beta \rightarrow \beta + \pi\,$.

\subsubsection{Massive IIA}

\begin{table}[t]
\renewcommand{\arraystretch}{1.5}
\begin{tabular}{!{\vrule width 1.5pt}c!{\vrule width 1pt}cccc!{\vrule width 1.5pt}}
\noalign{\hrule height 1.5pt}
$\,\, \mathcal{Q} \,\,$                   &  $\,\,\textrm{ISO}(7)\,\,$ & $\,\,\textrm{ISO}(6,1)\,\,$ & $\,\,\textrm{ISO}(5,2)\,\,$ & $\,\,\textrm{ISO}(4,3)\,\,$  \\
\noalign{\hrule height 1.5pt}
$m^{-2/3} \, g^{5/3} \, \mathfrak{p}^0$ & $\phantom{-}\frac{1}{6}$  & $\phantom{-}\frac{1}{14}$ & $-\frac{1}{14}$ & $-\frac{1}{6}$  \\
$g \, p^1$                                               &  $-\frac{1}{3}$ & $-\frac{1}{7}$ & $-\frac{1}{7}$ &  $-\frac{1}{3}$  \\
$g \, p^2$                                               &  $-\frac{1}{3}$ & $-\frac{1}{7}$ & $-\frac{1}{7}$ & $-\frac{1}{3}$   \\
$g \, p^3$                                               &  $-\frac{1}{3}$ & $-\frac{5}{7}$ & $\phantom{-}\frac{5}{7}$ &  $\phantom{-}\frac{1}{3}$  \\
\noalign{\hrule height 1.0pt}
$m^{1/3} \, g^{2/3} \, \mathfrak{e}_0$  &  $-\frac{1}{6}$ & $\phantom{-}\frac{1}{14}$  & $\phantom{-}\frac{1}{14}$ & $-\frac{1}{6}$ \\
$m^{-1/3} \, g^{4/3} \, e_1$                  &  $-\frac{1}{6}$ & $-\frac{3}{14}$ & $\phantom{-}\frac{3}{14}$ & $\phantom{-}\frac{1}{6}$   \\
$m^{-1/3} \, g^{4/3} \, e_2$                  &  $-\frac{1}{6}$ & $-\frac{3}{14}$ & $\phantom{-}\frac{3}{14}$ &  $\phantom{-}\frac{1}{6}$  \\
$m^{-1/3} \, g^{4/3} \, e_3$                  &  $-\frac{1}{6}$ & $-\frac{3}{14}$ & $-\frac{3}{14}$ & $-\frac{1}{6}$   \\
\noalign{\hrule height 1.5pt}
$\beta$ & $\frac{\pi}{6}$ &  $-\frac{\pi}{6}$ &  $-\frac{5\pi}{6}$  &  $\frac{5\pi}{6}$ \\
\noalign{\hrule height 1.5pt}
\end{tabular}
\caption{Vector charges supporting supersymmetric $\,\textrm{AdS}_{2} \times \Sigma_{2}\,$ solutions in the massive IIA models.}
\label{Table:Charges_AdS2_IIA}
\end{table}

Quarter-BPS $\,\textrm{AdS}_{2} \times \Sigma_{2}\,$ solutions exist in all the $\,\textrm{ISO}(p,q)\,$ models for different values of the vector of charges $\,\mathcal{Q}\,$ (see \mbox{Table~\ref{Table:Charges_AdS2_IIA}}). The solutions have non-zero magnetic $\,\mathfrak{p}^{\Lambda}\,$ and electric $\,\mathfrak{e}_{\Lambda}\,$ charges in (\ref{charges_p_e}), and require  $\,|\vec{\zeta}|^2=0\,$ and a non-trivial value of the dilaton $\,e^{\phi}\,$ in the universal \mbox{hypermultiplet}. 

For the $\,\textrm{ISO}(7)\,$ and $\,\textrm{ISO}(4,3)\,$ models the scalars $\,z^{i}\,$ are fixed at the values in (\ref{AdS4_ISO7_N2}) and (\ref{AdS4_ISO43_N2}), respectively, whereas the dilaton takes the value in (\ref{phi_AdS4_ISO7_ISO43_N2}). In both cases the solution has a hyperbolic horizon $\,{\Sigma_{2}=\mathbb{H}^2}\,$ ($\kappa=-1$) and radii given by
\begin{equation}
\label{LAdS_2_ISO7_ISO43}
L^2_{\textrm{AdS}_{2}} = \frac{1}{4 \sqrt{3}} \, g^{-\frac{7}{3}}\, m^{\frac{1}{3}}   
\hspace{3mm} , \hspace{3mm}  
L^2_{\mathbb{H}^2} = \frac{1}{2 \sqrt{3}} \, \, g^{-\frac{7}{3}}\, m^{\frac{1}{3}} \ .
\end{equation}
For the $\,\textrm{ISO}(5,2)\,$ and $\,\textrm{ISO}(6,1)\,$ models the scalars $\,z^{i}\,$ are also fixed at the values in (\ref{AdS4_ISO7_N2}) and (\ref{AdS4_ISO43_N2}), respectively. However the dilaton in the universal hypermultiplet is fixed at the value
\begin{equation}
\label{AdS2_ISO52_ISO61}
e^{\phi} = \frac{\sqrt{2}}{\sqrt{3}} \,  \left(\frac{g}{m}\right)^{\frac{1}{3}}  \ .
\end{equation}
The solutions have a spherical horizon $\,\Sigma_{2}=\mathbb{S}^2\,$ (${\kappa=+1}$) and radii given by
\begin{equation}
\label{LAdS_2_ISO52_ISO61}
L^2_{\textrm{AdS}_{2}} = \frac{3 \sqrt{3}}{4} \, g^{-\frac{7}{3}}\, m^{\frac{1}{3}}
\hspace{3mm} , \hspace{3mm} 
 L^2_{\mathbb{S}^2} =\frac{3 \sqrt{3}}{14}  \, \, g^{-\frac{7}{3}}\, m^{\frac{1}{3}} \ .
\end{equation}

Once again, a set of equivalent solutions is obtained upon replacing $\,\mathcal{Q} \rightarrow -\mathcal{Q}\,$ and $\,\beta \rightarrow \beta + \pi\,$.

\subsection{AdS$_{4}$ black hole solutions}

Extremal R--N black holes interpolating between the (charged version \cite{Hristov:2011ye} of) $\,\textrm{AdS}_{4}\,$ and $\,{\textrm{AdS}_{2} \times \Sigma_{2}}\,$ solutions previously found can be constructed. These black holes have constant scalars and can be viewed as non-minimal M-theory and massive~IIA incarnations of the universal black hole in (\ref{metric_universal_BH})-(\ref{metric_function_universal_BH}) with a hyperbolic horizon. 

In the context of M-theory reduced on $\,\textrm{H}^{(p,q)}\,$ spaces, two versions of such a black hole occur in each of the $\,\textrm{SO}(8)\,$ and $\,\textrm{SO}(4,4)\,$ models. The first one involves a trivial configuration of the universal hypermultiplet and interpolates between an AdS$_{4}$ geometry with radius (\ref{LAdS4_SO8SO44_N8}) in the ultraviolet (UV at $\,r \rightarrow \infty\,$) and an $\,\textrm{AdS}_{2} \times \mathbb{H}^{2}\,$ geometry with radii (\ref{LAdS_2_SO8_SO44_N8}) in the infrared (IR at $\,r \rightarrow r_{h}\,$). The case of the SO(8) model arising from $\,\textrm{H}^{(8,0)}=\textrm{S}^7\,$ has been extensively studied in the literature, see \textit{e.g.} \cite{Caldarelli:1998hg,Cacciatori:2009iz,DallAgata:2010ejj}, also from a holographic perspective \cite{Benini:2015eyy,Benini:2016rke,Cabo-Bizet:2017jsl,Azzurli:2017kxo}. The second version involves a non-trivial configuration of the universal hypermultiplet (\ref{AdS4_SOpq_N2_hyper}) and interpolates between an AdS$_{4}$ geometry with radius (\ref{LAdS4_SO8SO44_N2}) in the UV and an $\,\textrm{AdS}_{2} \times \mathbb{H}^{2}\,$ geometry with radii (\ref{LAdS_2_SO8_SO44_N2}) in the IR\footnote{See \cite{Halmagyi:2013sla} for a similar model based on the $\,Q^{111}\,$ truncation of M-theory \cite{Cassani:2012pj}.}. For the SO(8) model, it would be interesting to perform a holographic counting of AdS$_{4}$ black hole microstates in the field theory context of \cite{Benna:2008zy}.

In the context of massive IIA reduced on $\,\textrm{H}^{(p,q)}\,$ spaces, there exists a universal black hole both in the $\,\textrm{ISO}(7)\,$ and $\,\textrm{ISO}(4,3)\,$ models. It involves a non-trivial value of the dilaton field in the universal hypermultiplet (\ref{phi_AdS4_ISO7_ISO43_N2}) and interpolates between an AdS$_{4}$ geometry with radius (\ref{LAdS_4_ISO7_ISO43}) in the UV and an $\,\textrm{AdS}_{2} \times \mathbb{H}^{2}\,$ geometry with radii (\ref{LAdS_2_ISO7_ISO43}) in the IR. In the case of the ISO(7) theory arising from $\,\textrm{H}^{(7,0)}=\textrm{S}^6\,$, such a black hole has recently been constructed in \cite{Guarino:2017eag,Guarino:2017pkw} and connected to a universal RG flow across dimensions in \cite{Azzurli:2017kxo} (see \cite{Hosseini:2017fjo,Benini:2017oxt}) using holography.

\section{Summary and final remarks}
\label{Sec:Conclusions}

In this note we have investigated supersymmetric $\,\textrm{AdS}_{4}\,$, $\,\textrm{AdS}_{2} \times \Sigma_{2}\,$ and universal AdS$_{4}$ black hole solutions in non-minimal $\,\mathcal{N}=2\,$ supergravity models with three vector multiplets (STU-model) and Abelian gaugings of the universal hypermultiplet. We have performed a systematic characterisation of $\,\mathcal{N}=2\,$ models that arise from 11D and massive IIA supergravity when reduced on $\,\textrm{H}^{(p,q)}\,$ spaces down to four dimension. More concretely, the models correspond to the $\,\textrm{U}(1)^2\,$ invariant sector of the $\,\textrm{SO}(p,q)\,$ (M-theory) and $\,\textrm{ISO}(p,q)\,$ (massive IIA) gauged maximal supergravities resulting upon reduction of the higher-dimensional theories. In M-theory models, the gaugings involve a duality-hidden symmetry pair of the universal hypermultiplet. In contrast, only duality symmetries of the universal hypermultiplet are gauged in massive IIA models. Supersymmetric solutions turn to populate different domains of the K\"ahler cone both in the M-theory and massive IIA cases. 

Future lines to explore are immediately envisaged. The first one is the higher-dimensional description of the models based on non-compact $\,\textrm{H}^{(p,q)}\,$ reductions. Solutions of such models often lie in a different domain of the K\"ahler cone than their counterparts based on sphere reductions. In addition, for the case of sphere reductions, it would also be interesting to have a higher-dimensional picture of the AdS$_{4}$ black holes involving a non-trivial universal hypermultiplet as the gravitational backreaction of bound states of \mbox{M-/D-}branes wrapping a Riemann surface \cite{Maldacena:2000mw,Gauntlett:2001qs}. To this end, connecting the four-dimensional fields and charges to higher-dimensional backgrounds of 11D and massive IIA supergravity is required. A suitable framework to obtain such uplifts is provided by the duality-covariant formulation of 11D \cite{Hohm:2013uia} and massive IIA supergravity \cite{Ciceri:2016dmd} in terms of an exceptional field theory. \mbox{Using} this framework, general uplifting formulae have been systematically derived for the consistent truncation of \mbox{M-theory} and massive IIA on spheres\footnote{See \cite{Varela:2015ywx} for a recent derivation of the complete $\,D=11\,$ embedding of $\,\textrm{SO}(8)\,$ gauged maximal supergravity using the method introduced in \cite{Guarino:2015vca} for the $\,D=10\,$ embedding of the $\,\textrm{ISO}(7)\,$ theory.} and hyperboloids \cite{Hohm:2014qga,Inverso:2016eet} down to a gauged maximal supergravity in four dimensions. It would then be interesting to uplift the $\,\mathcal{N}=2\,$ supergravity models constructed in this note to backgrounds of 11D and massive IIA supergravity involving $\,\textrm{H}^{(p,q)}\,$ spaces.

Lastly, the $\,\mathcal{N}=2\,$ formulation of M-theory models presented in this note can be straightforwardly used to describe $\,\textrm{SO}(p,q)\,$ models with dyonic gaugings, the prototypical example being the \mbox{$\omega$-deformed} version of the $\,\textrm{SO}(8)\,$ theory \cite{Dall'Agata:2012bb}. In this case, and unlike for the original STU-model with only vector multiplets \cite{Inverso:2015viq}, the presence of the universal hypermultiplet makes the \mbox{Lagrangian} in (\ref{Lagrangian_N2}) sensitive to the electric-magnetic deformation parameter $\,\omega\,$ and affects the structure of supersymmetric solutions. For instance, two inequivalent $\,\mathcal{N} = 2\,$ AdS$_{4}$ solutions preserving an $\,\textrm{SU}(3) \times \textrm{U}(1)\,$ symmetry in the full theory appear at generic values of $\,\omega\,$ \cite{Borghese:2012zs}. It is therefore interesting to understand the structure of AdS$_{4}$ black hole solutions in this setup \cite{Guarino:wip} for which a higher-dimensional description remains \mbox{elusive}~\cite{Lee:2015xga}. We hope to come back to these and related issues in the near future.

\vspace{2mm}
\noindent{\bf Acknowledgements:} This work is partially supported by the ERC Advanced Grant ``High-Spin-Grav" and by F.R.S.-FNRS through the conventions PDRT.1025.14 and IISN-4.4503.15.

\section*{Appendix: Embedding in maximal supergravity}
\label{sec:appendix}

In this appendix we provide the embedding of the non-minimal $\,\mathcal{N}=2\,$ supergravity models analysed in this note into $\,\mathcal{N}=8\,$ (maximal) supergravity \cite{Cremmer:1979up}. In its ungauged version, maximal supergravity possesses an E$_{7(7)}$ global symmetry group. This group plays a central role in systematically constructing the gauged versions of the theory using the embedding tensor formalism \cite{deWit:2007mt}: in a gauged maximal supergravity, a specific subgroup of E$_{7(7)}$ is promoted from global to local in what is known as the gauging procedure. In this note we focus on $\,\textrm{SO}(p,q)\,$ and $\,\textrm{ISO}(p,q)\,$ subgroups of $\, \textrm{SL}(8) \subset \textrm{E}_{7(7)}$. These gauged supergravities appear upon reduction of 11D and \mbox{massive IIA} supergravity on $\,\textrm{H}^{(p,q)}\,$ spaces.

Let us start by introducing a fundamental SL(8) index $\,A=1,...8\,$. In the SL(8) basis, the E$_{7(7)}$ generators $\,t_{\alpha=1,...,133}\,$ have a decomposition $\,\textbf{133} \rightarrow \textbf{63} + \textbf{70}\,$. These are the $\,\textbf{63}\,$ generators $\,{t_{A}}^{B}\,$ of SL(8), with vanishing trace $\,{t_{A}}^{A}=0\,$, together with $\,\textbf{70}\,$ generators $\,t_{ABCD}=t_{[ABCD]}\,$. The fundamental representation of E$_{7(7)}$ decomposes as $\,\textbf{56} \rightarrow \textbf{28} + \textbf{28'}\,$, which translates into a splitting of the E$_{7(7)}$ fundamental index of the form $\,_\mathbb{M} \rightarrow _{[AB]} \oplus ^{[AB]}\,$. The entries of the $\,56 \times 56\,$ matrices $\,{[t_{\alpha}]_{\mathbb{M}}}^{\mathbb{N}\,}$ are given by
\begin{equation}
\label{63Gener}
\begin{array}{llll}
{[{t_{A}}^{B}]_{[CD]}}^{[EF]} &=& 4 \, \left( \delta_{[C}^{B} \, \delta_{D]A}^{EF} + \frac{1}{8} \, \delta_{A}^{B} \, \delta_{CD}^{EF}  \right) & ,\\[2mm]
{[{t_{A}}^{B}]^{[EF]}}_{[CD]} &=& - {[{t_{A}}^{B}]_{[CD]}}^{[EF]} & ,
\end{array}
\end{equation}
for the SL(8) generators $\,{t_{A}}^{B}\,$. The generators $\,t_{ABCD}\,$ completing SL(8) to E$_{7(7)}$ take the form
\begin{equation}
\label{70Gener}
\begin{array}{llll}
[t_{ABCD}]_{[EF][GH]} &=& \frac{2}{4!} \, \epsilon_{ABCDEFGH} & , \\[2mm]
[t_{ABCD}]^{[EF][GH]} &=& 2 \, \delta_{ABCD}^{EFGH} & .
\end{array}
\end{equation}
The electric $\,\textrm{SO}(p,q)\,$ and dyonic $\,\textrm{ISO}(p,q)\,$ gaugings of maximal supergravity belong to $\,\textrm{SL}(8)\subset \textrm{E}_{7(7)}\,$ and are specified by an embedding tensor \cite{deWit:2007mt} of the form 
\begin{equation}
\Theta_{AB}{}^{C}{}_{D} = 2 \,  \delta_{[A}^{C} \, \eta_{B]D}
\hspace{3mm} , \hspace{3mm}
\Theta^{AB}{}^{C}{}_{D} = 2 \,  \delta^{[A}_{D} \,  \tilde{\eta}^{B]C}  \ .
\end{equation}
The matrices $\,\eta_{AB}\,$ and $\,\tilde{\eta}^{AB}\,$  associated with the different gaugings are collected in Table~\ref{Table:eta_matrices}.

\begin{table}[t!]
\renewcommand{\arraystretch}{1.5}
\begin{tabular}{!{\vrule width 1.5pt}c!{\vrule width 1pt}c!{\vrule width 1pt}c!{\vrule width 1.5pt}}
\noalign{\hrule height 1.5pt}
 \,\,\textsc{gauging} \,\,      &  $\,\,\eta_{AB}\,\,$ & $\,\,\tilde{\eta}^{AB}\,\,$  \\
\noalign{\hrule height 1pt}
 $\textrm{SO}(8)$   &   $\textrm{diag}(+1,\mathbb{I}_{2},\mathbb{I}_{2},\mathbb{I}_{2},+1)$ & $0$   \\
 $\textrm{SO}(7,1)$   &   $\textrm{diag}(+1,\mathbb{I}_{2},\mathbb{I}_{2},\mathbb{I}_{2},-1)$ & $0$   \\
$\textrm{SO}(6,2)_{a}$   &    $\textrm{diag}(-1,\mathbb{I}_{2},\mathbb{I}_{2},\mathbb{I}_{2},-1)$ & $0$   \\
$\textrm{SO}(6,2)_{b}$   &    $\,\, \textrm{diag}(+1,\mathbb{I}_{2},\mathbb{I}_{2},-\mathbb{I}_{2},+1)\,\,$ & $0$   \\
 $\textrm{SO}(5,3)$   &   $\textrm{diag}(+1,\mathbb{I}_{2},\mathbb{I}_{2},-\mathbb{I}_{2},-1)$ & $0$   \\
 $\textrm{SO}(4,4)$   &    $\textrm{diag}(-1,\mathbb{I}_{2},\mathbb{I}_{2},-\mathbb{I}_{2},-1)$ & $0$   \\
\noalign{\hrule height 1pt}
 $\textrm{ISO}(7)$   &    $\textrm{diag}(+1,\mathbb{I}_{2},\mathbb{I}_{2},\mathbb{I}_{2},0)$ & $\,\,\textrm{diag}(0_{7},-m)\,\,$   \\
 $\textrm{ISO}(6,1)$   &    $\textrm{diag}(-1,\mathbb{I}_{2},\mathbb{I}_{2},\mathbb{I}_{2},0)$ & $\,\,\textrm{diag}(0_{7},-m)\,\,$   \\
 $\textrm{ISO}(5,2)$   &    $\textrm{diag}(+1,\mathbb{I}_{2},\mathbb{I}_{2},-\mathbb{I}_{2},0)$ & $\,\,\textrm{diag}(0_{7},-m)\,\,$   \\
 $\textrm{ISO}(4,3)$   &    $\textrm{diag}(-1,\mathbb{I}_{2},\mathbb{I}_{2},-\mathbb{I}_{2},0)$ & $\,\,\textrm{diag}(0_{7},-m)\,\,$   \\
\noalign{\hrule height 1.5pt}
\end{tabular}
\caption{Matrices $\,\eta_{AB}\,$ and $\,\tilde{\eta}^{AB}\,$ specifying the $\,\textrm{SO}(p,q)\,$ and $\,\textrm{ISO}(p,q)\,$ gaugings of maximal \mbox{supergravity}.}
\label{Table:eta_matrices}
\end{table}

The scalar sector of maximal supergravity consists of $\,70\,$ fields spanning a coset space $\,\textrm{E}_{7(7)}/\textrm{SU}(8)\,$. However, in this note we concentrate on non-minimal $\,\mathcal{N}=2\,$ supergravity models associated with the $\,\textrm{U}(1)^2\,$ invariant sector of the theory. The scalars in this sector span an $\,[\textrm{SL}(2)/\textrm{SO}(2)]^3 \, \times \, \textrm{SU}(2,1)/(\textrm{SU}(2)\times\textrm{U}(1))\,$ coset space associated with the following E$_{7(7)}$ generators. The $\,[\textrm{SL}(2)/\textrm{SO}(2)]^3\,$ factor is associated with
\begin{equation}
\label{Gener_SL2}
\begin{array}{ccll}
H_{\varphi_1}  & = &  {t_{4}}^{4} + {t_{5}}^{5} + {t_{6}}^{6} +  {t_{7}}^{7} - {t_{1}}^{1} - {t_{8}}^{8} - {t_{2}}^{2} - {t_{3}}^{3} & , \\[2mm]
H_{\varphi_2}  & = &  {t_{2}}^{2} + {t_{3}}^{3} + {t_{6}}^{6} +  {t_{7}}^{7} - {t_{1}}^{1} - {t_{8}}^{8} - {t_{4}}^{4} - {t_{5}}^{5} & , \\[2mm]
H_{\varphi_3}  & = &  {t_{2}}^{2} + {t_{3}}^{3} + {t_{4}}^{4} +  {t_{5}}^{5} - {t_{1}}^{1} - {t_{8}}^{8} - {t_{6}}^{6} - {t_{7}}^{7} & , \\[2mm]
g_{\chi_1} &=&  t_{1238} 
\hspace{3mm} , \hspace{3mm}
g_{\chi_2}  \,\,\,=\,\,\,  t_{1458}  
\hspace{3mm} , \hspace{3mm}
g_{\chi_3} \,\,\,=\,\,\,  t_{1678}  & .
\end{array}
\end{equation}
The $\,\textrm{SU}(2,1)/(\textrm{SU}(2)\times\textrm{U}(1))\,$ factor is associated with
\begin{equation}
\label{Gener_SU3}
\begin{array}{ccll}

H_{\phi} &=& \tfrac{1}{2} \, ({t_{8}}^{8}-{t_{1}}^{1}) 
\hspace{3mm} , \hspace{2mm}
g_{\sigma} \,\,\, = \,\,\,  {t_{8}}^{1} & , \\[2mm]

g_{\zeta} & = &   t_{8357}   - t_{8346} - t_{8256} - t_{8247}  & ,  \\[2mm]

g_{\tilde{\zeta}}   & = &   t_{8246} - t_{8257} - t_{8347} - t_{8356}  & .
\end{array}
\end{equation}
Using the above generators, the coset representative ${\mathcal{V}=\mathcal{V}_{\textrm{SK}} \times \mathcal{V}_{\textrm{QK}}}$ is obtained upon the exponentiations
\begin{equation}
\label{coset_SU3}
\begin{array}{llll}
\mathcal{V}_{\textrm{SK}}  &=& \prod_{i} e^{-12 \, \chi_{i} \, g_{\chi_{i}}} \,\,\, e^{\frac{1}{4} \, \varphi_{i} \, H_{\varphi_{i}}} & ,  \\[2mm]
\mathcal{V}_{\textrm{QK}}  &=&  e^{\sigma \,  g_{\sigma}} \,\,\, e^{-6 \,(\zeta \, g_{\zeta} \,+\, \tilde{\zeta} \, g_{\tilde{\zeta}}) }  \,\,\, e^{-2 \, \phi \, H_{\phi}} & . 
\end{array}
\end{equation}
Starting from the representative $\,\mathcal{V}\in \textrm{E}_{7(7)}/\textrm{SU}(8)\, $, the scalar matrix $\,\mathcal{M}_{\mathbb{MN}}\,$ entering the Lagrangian of $\,{\mathcal{N}=8}$ supergravity \cite{deWit:2007mt} is obtained as $\,{\mathcal{M}=\mathcal{V} \, \mathcal{V}^{t}}\,$. Plugging this matrix $\,\mathcal{M}\,$ into the kinetic terms of the maximal theory, $\,{e^{-1}  \mathcal{L}_{\textrm{kin}}=\frac{1}{96} \textrm{Tr} [ D_{\mu} \mathcal{M} \, D^{\mu}\mathcal{M}^{-1}]}\,$, we obtain the kinetic terms for the $\,\mathcal{N}=2\,$ models in (\ref{dSK}) and (\ref{dsQK_1}) upon identifying $\,{z^{i}=-\chi_{i}+ i \, e^{-\varphi_{i}}}\,$. Note that this parameterisation of the $\,\textrm{U}(1)^2\,$ invariant scalar coset in maximal supergravity is not compatible with configurations for which $\,\textrm{Im}z^i <0\,$. The scalar potential and covariant derivatives of the $\,\mathcal{N}=2\,$ models can be obtained from the ones of the maximal theory upon identifying the electric vectors $\,{\mathcal{A}^{AB}=\mathcal{A}^{[AB]}}\,$ in the maximal theory as $\,\{\mathcal{A}^0,\mathcal{A}^1,\mathcal{A}^2,\mathcal{A}^3 \} \equiv \{ \mathcal{A}^{18} , \mathcal{A}^{23} , \mathcal{A}^{45} , \mathcal{A}^{67} \}\,$ (and similarly for the magnetic vectors $\,{\tilde{\mathcal{A}}_{AB}}\,$), and then truncating away the non-singlet fields. We have verified that the $\,\mathcal{N}=8\,$ results precisely match the $\,\mathcal{N}=2\,$ results obtained from (\ref{VN2}) and (\ref{Dq}).

\bibliography{references}

\begin{thebibliography}{64}%
\makeatletter
\providecommand \@ifxundefined [1]{%
 \@ifx{#1\undefined}
}%
\providecommand \@ifnum [1]{%
 \ifnum #1\expandafter \@firstoftwo
 \else \expandafter \@secondoftwo
 \fi
}%
\providecommand \@ifx [1]{%
 \ifx #1\expandafter \@firstoftwo
 \else \expandafter \@secondoftwo
 \fi
}%
\providecommand \natexlab [1]{#1}%
\providecommand \enquote  [1]{``#1''}%
\providecommand \bibnamefont  [1]{#1}%
\providecommand \bibfnamefont [1]{#1}%
\providecommand \citenamefont [1]{#1}%
\providecommand \href@noop [0]{\@secondoftwo}%
\providecommand \href [0]{\begingroup \@sanitize@url \@href}%
\providecommand \@href[1]{\@@startlink{#1}\@@href}%
\providecommand \@@href[1]{\endgroup#1\@@endlink}%
\providecommand \@sanitize@url [0]{\catcode `\\12\catcode `\$12\catcode
  `\&12\catcode `\#12\catcode `\^12\catcode `\_12\catcode `\%12\relax}%
\providecommand \@@startlink[1]{}%
\providecommand \@@endlink[0]{}%
\providecommand \url  [0]{\begingroup\@sanitize@url \@url }%
\providecommand \@url [1]{\endgroup\@href {#1}{\urlprefix }}%
\providecommand \urlprefix  [0]{URL }%
\providecommand \Eprint [0]{\href }%
\providecommand \doibase [0]{http://dx.doi.org/}%
\providecommand \selectlanguage [0]{\@gobble}%
\providecommand \bibinfo  [0]{\@secondoftwo}%
\providecommand \bibfield  [0]{\@secondoftwo}%
\providecommand \translation [1]{[#1]}%
\providecommand \BibitemOpen [0]{}%
\providecommand \bibitemStop [0]{}%
\providecommand \bibitemNoStop [0]{.\EOS\space}%
\providecommand \EOS [0]{\spacefactor3000\relax}%
\providecommand \BibitemShut  [1]{\csname bibitem#1\endcsname}%
\let\auto@bib@innerbib\@empty
\bibitem [{\citenamefont {Azzurli}\ \emph {et~al.}(2017)\citenamefont
  {Azzurli}, \citenamefont {Bobev}, \citenamefont {Crichigno}, \citenamefont
  {Min},\ and\ \citenamefont {Zaffaroni}}]{Azzurli:2017kxo}%
  \BibitemOpen
  \bibfield  {author} {\bibinfo {author} {\bibfnamefont {F.}~\bibnamefont
  {Azzurli}}, \bibinfo {author} {\bibfnamefont {N.}~\bibnamefont {Bobev}},
  \bibinfo {author} {\bibfnamefont {P.~M.}\ \bibnamefont {Crichigno}}, \bibinfo
  {author} {\bibfnamefont {V.~S.}\ \bibnamefont {Min}}, \ and\ \bibinfo
  {author} {\bibfnamefont {A.}~\bibnamefont {Zaffaroni}},\ }\href@noop {} {\
  (\bibinfo {year} {2017})},\ \Eprint {http://arxiv.org/abs/1707.04257}
  {arXiv:1707.04257 [hep-th]} \BibitemShut {NoStop}%
\bibitem [{\citenamefont {Caldarelli}\ and\ \citenamefont
  {Klemm}(1999)}]{Caldarelli:1998hg}%
  \BibitemOpen
  \bibfield  {author} {\bibinfo {author} {\bibfnamefont {M.~M.}\ \bibnamefont
  {Caldarelli}}\ and\ \bibinfo {author} {\bibfnamefont {D.}~\bibnamefont
  {Klemm}},\ }\href {\doibase 10.1016/S0550-3213(98)00846-3} {\bibfield
  {journal} {\bibinfo  {journal} {Nucl. Phys.}\ }\textbf {\bibinfo {volume}
  {B545}},\ \bibinfo {pages} {434} (\bibinfo {year} {1999})},\ \Eprint
  {http://arxiv.org/abs/hep-th/9808097} {arXiv:hep-th/9808097 [hep-th]}
  \BibitemShut {NoStop}%
\bibitem [{\citenamefont {Freedman}\ and\ \citenamefont
  {Das}(1977)}]{Freedman:1976aw}%
  \BibitemOpen
  \bibfield  {author} {\bibinfo {author} {\bibfnamefont {D.~Z.}\ \bibnamefont
  {Freedman}}\ and\ \bibinfo {author} {\bibfnamefont {A.~K.}\ \bibnamefont
  {Das}},\ }\href {\doibase 10.1016/0550-3213(77)90041-4} {\bibfield  {journal}
  {\bibinfo  {journal} {Nucl. Phys.}\ }\textbf {\bibinfo {volume} {B120}},\
  \bibinfo {pages} {221} (\bibinfo {year} {1977})}\BibitemShut {NoStop}%
\bibitem [{\citenamefont {Romans}(1992)}]{Romans:1991nq}%
  \BibitemOpen
  \bibfield  {author} {\bibinfo {author} {\bibfnamefont {L.~J.}\ \bibnamefont
  {Romans}},\ }\href {\doibase 10.1016/0550-3213(92)90684-4} {\bibfield
  {journal} {\bibinfo  {journal} {Nucl. Phys.}\ }\textbf {\bibinfo {volume}
  {B383}},\ \bibinfo {pages} {395} (\bibinfo {year} {1992})},\ \Eprint
  {http://arxiv.org/abs/hep-th/9203018} {arXiv:hep-th/9203018 [hep-th]}
  \BibitemShut {NoStop}%
\bibitem [{\citenamefont {Bobev}\ and\ \citenamefont
  {Crichigno}(2017)}]{Bobev:2017uzs}%
  \BibitemOpen
  \bibfield  {author} {\bibinfo {author} {\bibfnamefont {N.}~\bibnamefont
  {Bobev}}\ and\ \bibinfo {author} {\bibfnamefont {P.~M.}\ \bibnamefont
  {Crichigno}},\ }\href {\doibase 10.1007/JHEP12(2017)065} {\bibfield
  {journal} {\bibinfo  {journal} {JHEP}\ }\textbf {\bibinfo {volume} {12}},\
  \bibinfo {pages} {065} (\bibinfo {year} {2017})},\ \Eprint
  {http://arxiv.org/abs/1708.05052} {arXiv:1708.05052 [hep-th]} \BibitemShut
  {NoStop}%
\bibitem [{\citenamefont {Cremmer}\ \emph {et~al.}(1978)\citenamefont
  {Cremmer}, \citenamefont {Julia},\ and\ \citenamefont
  {Scherk}}]{Cremmer:1978km}%
  \BibitemOpen
  \bibfield  {author} {\bibinfo {author} {\bibfnamefont {E.}~\bibnamefont
  {Cremmer}}, \bibinfo {author} {\bibfnamefont {B.}~\bibnamefont {Julia}}, \
  and\ \bibinfo {author} {\bibfnamefont {J.}~\bibnamefont {Scherk}},\ }\href
  {\doibase 10.1016/0370-2693(78)90894-8} {\bibfield  {journal} {\bibinfo
  {journal} {Phys.Lett.}\ }\textbf {\bibinfo {volume} {B76}},\ \bibinfo {pages}
  {409} (\bibinfo {year} {1978})}\BibitemShut {NoStop}%
\bibitem [{\citenamefont {Gauntlett}\ and\ \citenamefont
  {Varela}(2007)}]{Gauntlett:2007ma}%
  \BibitemOpen
  \bibfield  {author} {\bibinfo {author} {\bibfnamefont {J.~P.}\ \bibnamefont
  {Gauntlett}}\ and\ \bibinfo {author} {\bibfnamefont {O.}~\bibnamefont
  {Varela}},\ }\href {\doibase 10.1103/PhysRevD.76.126007} {\bibfield
  {journal} {\bibinfo  {journal} {Phys.Rev.}\ }\textbf {\bibinfo {volume}
  {D76}},\ \bibinfo {pages} {126007} (\bibinfo {year} {2007})},\ \Eprint
  {http://arxiv.org/abs/0707.2315} {arXiv:0707.2315 [hep-th]} \BibitemShut
  {NoStop}%
\bibitem [{\citenamefont {Romans}(1986)}]{Romans:1985tz}%
  \BibitemOpen
  \bibfield  {author} {\bibinfo {author} {\bibfnamefont {L.}~\bibnamefont
  {Romans}},\ }\href {\doibase 10.1016/0370-2693(86)90375-8} {\bibfield
  {journal} {\bibinfo  {journal} {Phys.Lett.}\ }\textbf {\bibinfo {volume}
  {B169}},\ \bibinfo {pages} {374} (\bibinfo {year} {1986})}\BibitemShut
  {NoStop}%
\bibitem [{\citenamefont {Aharony}\ \emph {et~al.}(2008)\citenamefont
  {Aharony}, \citenamefont {Bergman}, \citenamefont {Jafferis},\ and\
  \citenamefont {Maldacena}}]{Aharony:2008ug}%
  \BibitemOpen
  \bibfield  {author} {\bibinfo {author} {\bibfnamefont {O.}~\bibnamefont
  {Aharony}}, \bibinfo {author} {\bibfnamefont {O.}~\bibnamefont {Bergman}},
  \bibinfo {author} {\bibfnamefont {D.~L.}\ \bibnamefont {Jafferis}}, \ and\
  \bibinfo {author} {\bibfnamefont {J.}~\bibnamefont {Maldacena}},\ }\href
  {\doibase 10.1088/1126-6708/2008/10/091} {\bibfield  {journal} {\bibinfo
  {journal} {JHEP}\ }\textbf {\bibinfo {volume} {10}},\ \bibinfo {pages} {091}
  (\bibinfo {year} {2008})},\ \Eprint {http://arxiv.org/abs/0806.1218}
  {arXiv:0806.1218 [hep-th]} \BibitemShut {NoStop}%
\bibitem [{\citenamefont {Guarino}\ \emph {et~al.}(2015)\citenamefont
  {Guarino}, \citenamefont {Jafferis},\ and\ \citenamefont
  {Varela}}]{Guarino:2015jca}%
  \BibitemOpen
  \bibfield  {author} {\bibinfo {author} {\bibfnamefont {A.}~\bibnamefont
  {Guarino}}, \bibinfo {author} {\bibfnamefont {D.~L.}\ \bibnamefont
  {Jafferis}}, \ and\ \bibinfo {author} {\bibfnamefont {O.}~\bibnamefont
  {Varela}},\ }\href {\doibase 10.1103/PhysRevLett.115.091601} {\bibfield
  {journal} {\bibinfo  {journal} {Phys. Rev. Lett.}\ }\textbf {\bibinfo
  {volume} {115}},\ \bibinfo {pages} {091601} (\bibinfo {year} {2015})},\
  \Eprint {http://arxiv.org/abs/1504.08009} {arXiv:1504.08009 [hep-th]}
  \BibitemShut {NoStop}%
\bibitem [{\citenamefont {Fluder}\ and\ \citenamefont
  {Sparks}(2015)}]{Fluder:2015eoa}%
  \BibitemOpen
  \bibfield  {author} {\bibinfo {author} {\bibfnamefont {M.}~\bibnamefont
  {Fluder}}\ and\ \bibinfo {author} {\bibfnamefont {J.}~\bibnamefont
  {Sparks}},\ }\href@noop {} {\  (\bibinfo {year} {2015})},\ \Eprint
  {http://arxiv.org/abs/1507.05817} {arXiv:1507.05817 [hep-th]} \BibitemShut
  {NoStop}%
\bibitem [{\citenamefont {Benini}\ and\ \citenamefont
  {Zaffaroni}(2015)}]{Benini:2015noa}%
  \BibitemOpen
  \bibfield  {author} {\bibinfo {author} {\bibfnamefont {F.}~\bibnamefont
  {Benini}}\ and\ \bibinfo {author} {\bibfnamefont {A.}~\bibnamefont
  {Zaffaroni}},\ }\href {\doibase 10.1007/JHEP07(2015)127} {\bibfield
  {journal} {\bibinfo  {journal} {JHEP}\ }\textbf {\bibinfo {volume} {07}},\
  \bibinfo {pages} {127} (\bibinfo {year} {2015})},\ \Eprint
  {http://arxiv.org/abs/1504.03698} {arXiv:1504.03698 [hep-th]} \BibitemShut
  {NoStop}%
\bibitem [{\citenamefont {Benini}\ \emph
  {et~al.}(2016{\natexlab{a}})\citenamefont {Benini}, \citenamefont {Hristov},\
  and\ \citenamefont {Zaffaroni}}]{Benini:2015eyy}%
  \BibitemOpen
  \bibfield  {author} {\bibinfo {author} {\bibfnamefont {F.}~\bibnamefont
  {Benini}}, \bibinfo {author} {\bibfnamefont {K.}~\bibnamefont {Hristov}}, \
  and\ \bibinfo {author} {\bibfnamefont {A.}~\bibnamefont {Zaffaroni}},\ }\href
  {\doibase 10.1007/JHEP05(2016)054} {\bibfield  {journal} {\bibinfo  {journal}
  {JHEP}\ }\textbf {\bibinfo {volume} {05}},\ \bibinfo {pages} {054} (\bibinfo
  {year} {2016}{\natexlab{a}})},\ \Eprint {http://arxiv.org/abs/1511.04085}
  {arXiv:1511.04085 [hep-th]} \BibitemShut {NoStop}%
\bibitem [{\citenamefont {Benini}\ \emph
  {et~al.}(2016{\natexlab{b}})\citenamefont {Benini}, \citenamefont {Hristov},\
  and\ \citenamefont {Zaffaroni}}]{Benini:2016rke}%
  \BibitemOpen
  \bibfield  {author} {\bibinfo {author} {\bibfnamefont {F.}~\bibnamefont
  {Benini}}, \bibinfo {author} {\bibfnamefont {K.}~\bibnamefont {Hristov}}, \
  and\ \bibinfo {author} {\bibfnamefont {A.}~\bibnamefont {Zaffaroni}},\
  }\href@noop {} {\  (\bibinfo {year} {2016}{\natexlab{b}})},\ \Eprint
  {http://arxiv.org/abs/1608.07294} {arXiv:1608.07294 [hep-th]} \BibitemShut
  {NoStop}%
\bibitem [{\citenamefont {Cabo-Bizet}\ \emph {et~al.}(2017)\citenamefont
  {Cabo-Bizet}, \citenamefont {Giraldo-Rivera},\ and\ \citenamefont
  {Pando~Zayas}}]{Cabo-Bizet:2017jsl}%
  \BibitemOpen
  \bibfield  {author} {\bibinfo {author} {\bibfnamefont {A.}~\bibnamefont
  {Cabo-Bizet}}, \bibinfo {author} {\bibfnamefont {V.~I.}\ \bibnamefont
  {Giraldo-Rivera}}, \ and\ \bibinfo {author} {\bibfnamefont {L.~A.}\
  \bibnamefont {Pando~Zayas}},\ }\href@noop {} {\  (\bibinfo {year} {2017})},\
  \Eprint {http://arxiv.org/abs/1701.07893} {arXiv:1701.07893 [hep-th]}
  \BibitemShut {NoStop}%
\bibitem [{\citenamefont {Hosseini}\ \emph {et~al.}(2017)\citenamefont
  {Hosseini}, \citenamefont {Hristov},\ and\ \citenamefont
  {Passias}}]{Hosseini:2017fjo}%
  \BibitemOpen
  \bibfield  {author} {\bibinfo {author} {\bibfnamefont {S.~M.}\ \bibnamefont
  {Hosseini}}, \bibinfo {author} {\bibfnamefont {K.}~\bibnamefont {Hristov}}, \
  and\ \bibinfo {author} {\bibfnamefont {A.}~\bibnamefont {Passias}},\ }\href
  {\doibase 10.1007/JHEP10(2017)190} {\bibfield  {journal} {\bibinfo  {journal}
  {JHEP}\ }\textbf {\bibinfo {volume} {10}},\ \bibinfo {pages} {190} (\bibinfo
  {year} {2017})},\ \Eprint {http://arxiv.org/abs/1707.06884} {arXiv:1707.06884
  [hep-th]} \BibitemShut {NoStop}%
\bibitem [{\citenamefont {Benini}\ \emph {et~al.}(2017)\citenamefont {Benini},
  \citenamefont {Khachatryan},\ and\ \citenamefont {Milan}}]{Benini:2017oxt}%
  \BibitemOpen
  \bibfield  {author} {\bibinfo {author} {\bibfnamefont {F.}~\bibnamefont
  {Benini}}, \bibinfo {author} {\bibfnamefont {H.}~\bibnamefont {Khachatryan}},
  \ and\ \bibinfo {author} {\bibfnamefont {P.}~\bibnamefont {Milan}},\
  }\href@noop {} {\  (\bibinfo {year} {2017})},\ \Eprint
  {http://arxiv.org/abs/1707.06886} {arXiv:1707.06886 [hep-th]} \BibitemShut
  {NoStop}%
\bibitem [{\citenamefont {de~Wit}\ and\ \citenamefont
  {Nicolai}(1987)}]{deWit:1986oxb}%
  \BibitemOpen
  \bibfield  {author} {\bibinfo {author} {\bibfnamefont {B.}~\bibnamefont
  {de~Wit}}\ and\ \bibinfo {author} {\bibfnamefont {H.}~\bibnamefont
  {Nicolai}},\ }\href {\doibase 10.1016/0550-3213(87)90253-7} {\bibfield
  {journal} {\bibinfo  {journal} {Nucl. Phys.}\ }\textbf {\bibinfo {volume}
  {B281}},\ \bibinfo {pages} {211} (\bibinfo {year} {1987})}\BibitemShut
  {NoStop}%
\bibitem [{\citenamefont {de~Wit}\ and\ \citenamefont
  {Nicolai}(1982)}]{deWit:1982ig}%
  \BibitemOpen
  \bibfield  {author} {\bibinfo {author} {\bibfnamefont {B.}~\bibnamefont
  {de~Wit}}\ and\ \bibinfo {author} {\bibfnamefont {H.}~\bibnamefont
  {Nicolai}},\ }\href {\doibase 10.1016/0550-3213(82)90120-1} {\bibfield
  {journal} {\bibinfo  {journal} {Nucl.Phys.}\ }\textbf {\bibinfo {volume}
  {B208}},\ \bibinfo {pages} {323} (\bibinfo {year} {1982})}\BibitemShut
  {NoStop}%
\bibitem [{\citenamefont {Duff}\ and\ \citenamefont {Liu}(1999)}]{Duff:1999gh}%
  \BibitemOpen
  \bibfield  {author} {\bibinfo {author} {\bibfnamefont {M.~J.}\ \bibnamefont
  {Duff}}\ and\ \bibinfo {author} {\bibfnamefont {J.~T.}\ \bibnamefont {Liu}},\
  }\href {\doibase 10.1016/S0550-3213(99)00299-0} {\bibfield  {journal}
  {\bibinfo  {journal} {Nucl. Phys.}\ }\textbf {\bibinfo {volume} {B554}},\
  \bibinfo {pages} {237} (\bibinfo {year} {1999})},\ \Eprint
  {http://arxiv.org/abs/hep-th/9901149} {arXiv:hep-th/9901149 [hep-th]}
  \BibitemShut {NoStop}%
\bibitem [{\citenamefont {Cvetic}\ \emph {et~al.}(1999)\citenamefont {Cvetic},
  \citenamefont {Duff}, \citenamefont {Hoxha}, \citenamefont {Liu},
  \citenamefont {Lu}, \citenamefont {Lu}, \citenamefont {Martinez-Acosta},
  \citenamefont {Pope}, \citenamefont {Sati},\ and\ \citenamefont
  {Tran}}]{Cvetic:1999xp}%
  \BibitemOpen
  \bibfield  {author} {\bibinfo {author} {\bibfnamefont {M.}~\bibnamefont
  {Cvetic}}, \bibinfo {author} {\bibfnamefont {M.~J.}\ \bibnamefont {Duff}},
  \bibinfo {author} {\bibfnamefont {P.}~\bibnamefont {Hoxha}}, \bibinfo
  {author} {\bibfnamefont {J.~T.}\ \bibnamefont {Liu}}, \bibinfo {author}
  {\bibfnamefont {H.}~\bibnamefont {Lu}}, \bibinfo {author} {\bibfnamefont
  {J.~X.}\ \bibnamefont {Lu}}, \bibinfo {author} {\bibfnamefont
  {R.}~\bibnamefont {Martinez-Acosta}}, \bibinfo {author} {\bibfnamefont
  {C.~N.}\ \bibnamefont {Pope}}, \bibinfo {author} {\bibfnamefont
  {H.}~\bibnamefont {Sati}}, \ and\ \bibinfo {author} {\bibfnamefont {T.~A.}\
  \bibnamefont {Tran}},\ }\href {\doibase 10.1016/S0550-3213(99)00419-8}
  {\bibfield  {journal} {\bibinfo  {journal} {Nucl. Phys.}\ }\textbf {\bibinfo
  {volume} {B558}},\ \bibinfo {pages} {96} (\bibinfo {year} {1999})},\ \Eprint
  {http://arxiv.org/abs/hep-th/9903214} {arXiv:hep-th/9903214 [hep-th]}
  \BibitemShut {NoStop}%
\bibitem [{\citenamefont {Guarino}\ and\ \citenamefont
  {Varela}(2015)}]{Guarino:2015vca}%
  \BibitemOpen
  \bibfield  {author} {\bibinfo {author} {\bibfnamefont {A.}~\bibnamefont
  {Guarino}}\ and\ \bibinfo {author} {\bibfnamefont {O.}~\bibnamefont
  {Varela}},\ }\href {\doibase 10.1007/JHEP12(2015)020} {\bibfield  {journal}
  {\bibinfo  {journal} {JHEP}\ }\textbf {\bibinfo {volume} {12}},\ \bibinfo
  {pages} {020} (\bibinfo {year} {2015})},\ \Eprint
  {http://arxiv.org/abs/1509.02526} {arXiv:1509.02526 [hep-th]} \BibitemShut
  {NoStop}%
\bibitem [{\citenamefont {Hull}(1984)}]{Hull:1984yy}%
  \BibitemOpen
  \bibfield  {author} {\bibinfo {author} {\bibfnamefont {C.}~\bibnamefont
  {Hull}},\ }\href {\doibase 10.1103/PhysRevD.30.760} {\bibfield  {journal}
  {\bibinfo  {journal} {Phys.Rev.}\ }\textbf {\bibinfo {volume} {D30}},\
  \bibinfo {pages} {760} (\bibinfo {year} {1984})}\BibitemShut {NoStop}%
\bibitem [{\citenamefont {Guarino}\ and\ \citenamefont
  {Varela}(2016)}]{Guarino:2015qaa}%
  \BibitemOpen
  \bibfield  {author} {\bibinfo {author} {\bibfnamefont {A.}~\bibnamefont
  {Guarino}}\ and\ \bibinfo {author} {\bibfnamefont {O.}~\bibnamefont
  {Varela}},\ }\href {\doibase 10.1007/JHEP02(2016)079} {\bibfield  {journal}
  {\bibinfo  {journal} {JHEP}\ }\textbf {\bibinfo {volume} {02}},\ \bibinfo
  {pages} {079} (\bibinfo {year} {2016})},\ \Eprint
  {http://arxiv.org/abs/1508.04432} {arXiv:1508.04432 [hep-th]} \BibitemShut
  {NoStop}%
\bibitem [{\citenamefont {Dall'Agata}\ \emph {et~al.}(2014)\citenamefont
  {Dall'Agata}, \citenamefont {Inverso},\ and\ \citenamefont
  {Marrani}}]{Dall'Agata:2014ita}%
  \BibitemOpen
  \bibfield  {author} {\bibinfo {author} {\bibfnamefont {G.}~\bibnamefont
  {Dall'Agata}}, \bibinfo {author} {\bibfnamefont {G.}~\bibnamefont {Inverso}},
  \ and\ \bibinfo {author} {\bibfnamefont {A.}~\bibnamefont {Marrani}},\ }\href
  {\doibase 10.1007/JHEP07(2014)133} {\bibfield  {journal} {\bibinfo  {journal}
  {JHEP}\ }\textbf {\bibinfo {volume} {1407}},\ \bibinfo {pages} {133}
  (\bibinfo {year} {2014})},\ \Eprint {http://arxiv.org/abs/1405.2437}
  {arXiv:1405.2437 [hep-th]} \BibitemShut {NoStop}%
\bibitem [{\citenamefont {Guarino}\ and\ \citenamefont
  {Tarrio}(2017)}]{Guarino:2017eag}%
  \BibitemOpen
  \bibfield  {author} {\bibinfo {author} {\bibfnamefont {A.}~\bibnamefont
  {Guarino}}\ and\ \bibinfo {author} {\bibfnamefont {J.}~\bibnamefont
  {Tarrio}},\ }\href {\doibase 10.1007/JHEP09(2017)141} {\bibfield  {journal}
  {\bibinfo  {journal} {JHEP}\ }\textbf {\bibinfo {volume} {09}},\ \bibinfo
  {pages} {141} (\bibinfo {year} {2017})},\ \Eprint
  {http://arxiv.org/abs/1703.10833} {arXiv:1703.10833 [hep-th]} \BibitemShut
  {NoStop}%
\bibitem [{\citenamefont {Cacciatori}\ and\ \citenamefont
  {Klemm}(2010)}]{Cacciatori:2009iz}%
  \BibitemOpen
  \bibfield  {author} {\bibinfo {author} {\bibfnamefont {S.~L.}\ \bibnamefont
  {Cacciatori}}\ and\ \bibinfo {author} {\bibfnamefont {D.}~\bibnamefont
  {Klemm}},\ }\href {\doibase 10.1007/JHEP01(2010)085} {\bibfield  {journal}
  {\bibinfo  {journal} {JHEP}\ }\textbf {\bibinfo {volume} {01}},\ \bibinfo
  {pages} {085} (\bibinfo {year} {2010})},\ \Eprint
  {http://arxiv.org/abs/0911.4926} {arXiv:0911.4926 [hep-th]} \BibitemShut
  {NoStop}%
\bibitem [{\citenamefont {Guarino}(2017)}]{Guarino:2017pkw}%
  \BibitemOpen
  \bibfield  {author} {\bibinfo {author} {\bibfnamefont {A.}~\bibnamefont
  {Guarino}},\ }\href {\doibase 10.1007/JHEP08(2017)100} {\bibfield  {journal}
  {\bibinfo  {journal} {JHEP}\ }\textbf {\bibinfo {volume} {08}},\ \bibinfo
  {pages} {100} (\bibinfo {year} {2017})},\ \Eprint
  {http://arxiv.org/abs/1706.01823} {arXiv:1706.01823 [hep-th]} \BibitemShut
  {NoStop}%
\bibitem [{\citenamefont {Hull}\ and\ \citenamefont
  {Warner}(1988)}]{Hull:1988jw}%
  \BibitemOpen
  \bibfield  {author} {\bibinfo {author} {\bibfnamefont {C.}~\bibnamefont
  {Hull}}\ and\ \bibinfo {author} {\bibfnamefont {N.}~\bibnamefont {Warner}},\
  }\href {\doibase 10.1088/0264-9381/5/12/005} {\bibfield  {journal} {\bibinfo
  {journal} {Class.Quant.Grav.}\ }\textbf {\bibinfo {volume} {5}},\ \bibinfo
  {pages} {1517} (\bibinfo {year} {1988})}\BibitemShut {NoStop}%
\bibitem [{\citenamefont {Guarino}(2015)}]{Guarino:2015tja}%
  \BibitemOpen
  \bibfield  {author} {\bibinfo {author} {\bibfnamefont {A.}~\bibnamefont
  {Guarino}},\ }\href {\doibase 10.1016/j.nuclphysb.2015.09.016} {\bibfield
  {journal} {\bibinfo  {journal} {Nucl. Phys.}\ }\textbf {\bibinfo {volume}
  {B900}},\ \bibinfo {pages} {501} (\bibinfo {year} {2015})},\ \Eprint
  {http://arxiv.org/abs/1508.05055} {arXiv:1508.05055 [hep-th]} \BibitemShut
  {NoStop}%
\bibitem [{\citenamefont {Cassani}\ \emph {et~al.}(2016)\citenamefont
  {Cassani}, \citenamefont {de~Felice}, \citenamefont {Petrini}, \citenamefont
  {Strickland-Constable},\ and\ \citenamefont {Waldram}}]{Cassani:2016ncu}%
  \BibitemOpen
  \bibfield  {author} {\bibinfo {author} {\bibfnamefont {D.}~\bibnamefont
  {Cassani}}, \bibinfo {author} {\bibfnamefont {O.}~\bibnamefont {de~Felice}},
  \bibinfo {author} {\bibfnamefont {M.}~\bibnamefont {Petrini}}, \bibinfo
  {author} {\bibfnamefont {C.}~\bibnamefont {Strickland-Constable}}, \ and\
  \bibinfo {author} {\bibfnamefont {D.}~\bibnamefont {Waldram}},\ }\href
  {\doibase 10.1007/JHEP08(2016)074} {\bibfield  {journal} {\bibinfo  {journal}
  {JHEP}\ }\textbf {\bibinfo {volume} {08}},\ \bibinfo {pages} {074} (\bibinfo
  {year} {2016})},\ \Eprint {http://arxiv.org/abs/1605.00563} {arXiv:1605.00563
  [hep-th]} \BibitemShut {NoStop}%
\bibitem [{\citenamefont {Inverso}\ \emph {et~al.}(2017)\citenamefont
  {Inverso}, \citenamefont {Samtleben},\ and\ \citenamefont
  {Trigiante}}]{Inverso:2016eet}%
  \BibitemOpen
  \bibfield  {author} {\bibinfo {author} {\bibfnamefont {G.}~\bibnamefont
  {Inverso}}, \bibinfo {author} {\bibfnamefont {H.}~\bibnamefont {Samtleben}},
  \ and\ \bibinfo {author} {\bibfnamefont {M.}~\bibnamefont {Trigiante}},\
  }\href {\doibase 10.1103/PhysRevD.95.066020} {\bibfield  {journal} {\bibinfo
  {journal} {Phys. Rev.}\ }\textbf {\bibinfo {volume} {D95}},\ \bibinfo {pages}
  {066020} (\bibinfo {year} {2017})},\ \Eprint
  {http://arxiv.org/abs/1612.05123} {arXiv:1612.05123 [hep-th]} \BibitemShut
  {NoStop}%
\bibitem [{\citenamefont {de~Wit}\ \emph {et~al.}(2005)\citenamefont {de~Wit},
  \citenamefont {Samtleben},\ and\ \citenamefont {Trigiante}}]{deWit:2005ub}%
  \BibitemOpen
  \bibfield  {author} {\bibinfo {author} {\bibfnamefont {B.}~\bibnamefont
  {de~Wit}}, \bibinfo {author} {\bibfnamefont {H.}~\bibnamefont {Samtleben}}, \
  and\ \bibinfo {author} {\bibfnamefont {M.}~\bibnamefont {Trigiante}},\ }\href
  {\doibase 10.1088/1126-6708/2005/09/016} {\bibfield  {journal} {\bibinfo
  {journal} {JHEP}\ }\textbf {\bibinfo {volume} {0509}},\ \bibinfo {pages}
  {016} (\bibinfo {year} {2005})},\ \Eprint
  {http://arxiv.org/abs/hep-th/0507289} {arXiv:hep-th/0507289 [hep-th]}
  \BibitemShut {NoStop}%
\bibitem [{\citenamefont {Cecotti}\ \emph {et~al.}(1989)\citenamefont
  {Cecotti}, \citenamefont {Ferrara},\ and\ \citenamefont
  {Girardello}}]{Cecotti:1988qn}%
  \BibitemOpen
  \bibfield  {author} {\bibinfo {author} {\bibfnamefont {S.}~\bibnamefont
  {Cecotti}}, \bibinfo {author} {\bibfnamefont {S.}~\bibnamefont {Ferrara}}, \
  and\ \bibinfo {author} {\bibfnamefont {L.}~\bibnamefont {Girardello}},\
  }\href {\doibase 10.1142/S0217751X89000972} {\bibfield  {journal} {\bibinfo
  {journal} {Int. J. Mod. Phys.}\ }\textbf {\bibinfo {volume} {A4}},\ \bibinfo
  {pages} {2475} (\bibinfo {year} {1989})}\BibitemShut {NoStop}%
\bibitem [{\citenamefont {Klemm}\ \emph {et~al.}(2016)\citenamefont {Klemm},
  \citenamefont {Petri},\ and\ \citenamefont {Rabbiosi}}]{Klemm:2016wng}%
  \BibitemOpen
  \bibfield  {author} {\bibinfo {author} {\bibfnamefont {D.}~\bibnamefont
  {Klemm}}, \bibinfo {author} {\bibfnamefont {N.}~\bibnamefont {Petri}}, \ and\
  \bibinfo {author} {\bibfnamefont {M.}~\bibnamefont {Rabbiosi}},\ }\href
  {\doibase 10.1007/JHEP04(2016)008} {\bibfield  {journal} {\bibinfo  {journal}
  {JHEP}\ }\textbf {\bibinfo {volume} {04}},\ \bibinfo {pages} {008} (\bibinfo
  {year} {2016})},\ \Eprint {http://arxiv.org/abs/1602.01334} {arXiv:1602.01334
  [hep-th]} \BibitemShut {NoStop}%
\bibitem [{\citenamefont {Michelson}(1997)}]{Michelson:1996pn}%
  \BibitemOpen
  \bibfield  {author} {\bibinfo {author} {\bibfnamefont {J.}~\bibnamefont
  {Michelson}},\ }\href {\doibase 10.1016/S0550-3213(97)00184-3} {\bibfield
  {journal} {\bibinfo  {journal} {Nucl.Phys.}\ }\textbf {\bibinfo {volume}
  {B495}},\ \bibinfo {pages} {127} (\bibinfo {year} {1997})},\ \Eprint
  {http://arxiv.org/abs/hep-th/9610151} {arXiv:hep-th/9610151 [hep-th]}
  \BibitemShut {NoStop}%
\bibitem [{\citenamefont {de~Wit}\ and\ \citenamefont
  {Van~Proeyen}(1990)}]{deWit:1990na}%
  \BibitemOpen
  \bibfield  {author} {\bibinfo {author} {\bibfnamefont {B.}~\bibnamefont
  {de~Wit}}\ and\ \bibinfo {author} {\bibfnamefont {A.}~\bibnamefont
  {Van~Proeyen}},\ }\href {\doibase 10.1016/0370-2693(90)90864-3} {\bibfield
  {journal} {\bibinfo  {journal} {Phys. Lett.}\ }\textbf {\bibinfo {volume}
  {B252}},\ \bibinfo {pages} {221} (\bibinfo {year} {1990})}\BibitemShut
  {NoStop}%
\bibitem [{\citenamefont {de~Wit}\ \emph {et~al.}(1993)\citenamefont {de~Wit},
  \citenamefont {Vanderseypen},\ and\ \citenamefont
  {Van~Proeyen}}]{deWit:1992wf}%
  \BibitemOpen
  \bibfield  {author} {\bibinfo {author} {\bibfnamefont {B.}~\bibnamefont
  {de~Wit}}, \bibinfo {author} {\bibfnamefont {F.}~\bibnamefont
  {Vanderseypen}}, \ and\ \bibinfo {author} {\bibfnamefont {A.}~\bibnamefont
  {Van~Proeyen}},\ }\href {\doibase 10.1016/0550-3213(93)90413-J} {\bibfield
  {journal} {\bibinfo  {journal} {Nucl. Phys.}\ }\textbf {\bibinfo {volume}
  {B400}},\ \bibinfo {pages} {463} (\bibinfo {year} {1993})},\ \Eprint
  {http://arxiv.org/abs/hep-th/9210068} {arXiv:hep-th/9210068 [hep-th]}
  \BibitemShut {NoStop}%
\bibitem [{\citenamefont {de~Wit}\ and\ \citenamefont
  {Van~Proeyen}(1994)}]{deWit:1993rr}%
  \BibitemOpen
  \bibfield  {author} {\bibinfo {author} {\bibfnamefont {B.}~\bibnamefont
  {de~Wit}}\ and\ \bibinfo {author} {\bibfnamefont {A.}~\bibnamefont
  {Van~Proeyen}},\ }\href {\doibase 10.1142/S0218271894000058} {\bibfield
  {journal} {\bibinfo  {journal} {Int. J. Mod. Phys.}\ }\textbf {\bibinfo
  {volume} {D3}},\ \bibinfo {pages} {31} (\bibinfo {year} {1994})},\ \Eprint
  {http://arxiv.org/abs/hep-th/9310067} {arXiv:hep-th/9310067 [hep-th]}
  \BibitemShut {NoStop}%
\bibitem [{\citenamefont {Erbin}\ and\ \citenamefont
  {Halmagyi}(2015)}]{Erbin:2014hsa}%
  \BibitemOpen
  \bibfield  {author} {\bibinfo {author} {\bibfnamefont {H.}~\bibnamefont
  {Erbin}}\ and\ \bibinfo {author} {\bibfnamefont {N.}~\bibnamefont
  {Halmagyi}},\ }\href {\doibase 10.1007/JHEP05(2015)122} {\bibfield  {journal}
  {\bibinfo  {journal} {JHEP}\ }\textbf {\bibinfo {volume} {05}},\ \bibinfo
  {pages} {122} (\bibinfo {year} {2015})},\ \Eprint
  {http://arxiv.org/abs/1409.6310} {arXiv:1409.6310 [hep-th]} \BibitemShut
  {NoStop}%
\bibitem [{\citenamefont {Louis}\ \emph {et~al.}(2012)\citenamefont {Louis},
  \citenamefont {Smyth},\ and\ \citenamefont {Triendl}}]{Louis:2012ux}%
  \BibitemOpen
  \bibfield  {author} {\bibinfo {author} {\bibfnamefont {J.}~\bibnamefont
  {Louis}}, \bibinfo {author} {\bibfnamefont {P.}~\bibnamefont {Smyth}}, \ and\
  \bibinfo {author} {\bibfnamefont {H.}~\bibnamefont {Triendl}},\ }\href
  {\doibase 10.1007/JHEP08(2012)039} {\bibfield  {journal} {\bibinfo  {journal}
  {JHEP}\ }\textbf {\bibinfo {volume} {1208}},\ \bibinfo {pages} {039}
  (\bibinfo {year} {2012})},\ \Eprint {http://arxiv.org/abs/1204.3893}
  {arXiv:1204.3893 [hep-th]} \BibitemShut {NoStop}%
\bibitem [{\citenamefont {Freund}\ and\ \citenamefont
  {Rubin}(1980)}]{Freund:1980xh}%
  \BibitemOpen
  \bibfield  {author} {\bibinfo {author} {\bibfnamefont {P.~G.}\ \bibnamefont
  {Freund}}\ and\ \bibinfo {author} {\bibfnamefont {M.~A.}\ \bibnamefont
  {Rubin}},\ }\href {\doibase 10.1016/0370-2693(80)90590-0} {\bibfield
  {journal} {\bibinfo  {journal} {Phys.Lett.}\ }\textbf {\bibinfo {volume}
  {B97}},\ \bibinfo {pages} {233} (\bibinfo {year} {1980})}\BibitemShut
  {NoStop}%
\bibitem [{\citenamefont {Warner}(1983)}]{Warner:1983vz}%
  \BibitemOpen
  \bibfield  {author} {\bibinfo {author} {\bibfnamefont {N.}~\bibnamefont
  {Warner}},\ }\href {\doibase 10.1016/0370-2693(83)90383-0} {\bibfield
  {journal} {\bibinfo  {journal} {Phys.Lett.}\ }\textbf {\bibinfo {volume}
  {B128}},\ \bibinfo {pages} {169} (\bibinfo {year} {1983})}\BibitemShut
  {NoStop}%
\bibitem [{\citenamefont {Nicolai}\ and\ \citenamefont
  {Warner}(1985)}]{Nicolai:1985hs}%
  \BibitemOpen
  \bibfield  {author} {\bibinfo {author} {\bibfnamefont {H.}~\bibnamefont
  {Nicolai}}\ and\ \bibinfo {author} {\bibfnamefont {N.~P.}\ \bibnamefont
  {Warner}},\ }\href {\doibase 10.1016/0550-3213(85)90643-1} {\bibfield
  {journal} {\bibinfo  {journal} {Nucl. Phys.}\ }\textbf {\bibinfo {volume}
  {B259}},\ \bibinfo {pages} {412} (\bibinfo {year} {1985})}\BibitemShut
  {NoStop}%
\bibitem [{\citenamefont {Corrado}\ \emph {et~al.}(2002)\citenamefont
  {Corrado}, \citenamefont {Pilch},\ and\ \citenamefont
  {Warner}}]{Corrado:2001nv}%
  \BibitemOpen
  \bibfield  {author} {\bibinfo {author} {\bibfnamefont {R.}~\bibnamefont
  {Corrado}}, \bibinfo {author} {\bibfnamefont {K.}~\bibnamefont {Pilch}}, \
  and\ \bibinfo {author} {\bibfnamefont {N.~P.}\ \bibnamefont {Warner}},\
  }\href {\doibase 10.1016/S0550-3213(02)00134-7} {\bibfield  {journal}
  {\bibinfo  {journal} {Nucl. Phys.}\ }\textbf {\bibinfo {volume} {B629}},\
  \bibinfo {pages} {74} (\bibinfo {year} {2002})},\ \Eprint
  {http://arxiv.org/abs/hep-th/0107220} {arXiv:hep-th/0107220 [hep-th]}
  \BibitemShut {NoStop}%
\bibitem [{\citenamefont {Ahn}\ and\ \citenamefont {Itoh}(2002)}]{Ahn:2002eh}%
  \BibitemOpen
  \bibfield  {author} {\bibinfo {author} {\bibfnamefont {C.-h.}\ \bibnamefont
  {Ahn}}\ and\ \bibinfo {author} {\bibfnamefont {T.}~\bibnamefont {Itoh}},\
  }\href {\doibase 10.1016/S0550-3213(02)00871-4} {\bibfield  {journal}
  {\bibinfo  {journal} {Nucl. Phys.}\ }\textbf {\bibinfo {volume} {B646}},\
  \bibinfo {pages} {257} (\bibinfo {year} {2002})},\ \Eprint
  {http://arxiv.org/abs/hep-th/0208137} {arXiv:hep-th/0208137 [hep-th]}
  \BibitemShut {NoStop}%
\bibitem [{\citenamefont {Benna}\ \emph {et~al.}(2008)\citenamefont {Benna},
  \citenamefont {Klebanov}, \citenamefont {Klose},\ and\ \citenamefont
  {Smedback}}]{Benna:2008zy}%
  \BibitemOpen
  \bibfield  {author} {\bibinfo {author} {\bibfnamefont {M.}~\bibnamefont
  {Benna}}, \bibinfo {author} {\bibfnamefont {I.}~\bibnamefont {Klebanov}},
  \bibinfo {author} {\bibfnamefont {T.}~\bibnamefont {Klose}}, \ and\ \bibinfo
  {author} {\bibfnamefont {M.}~\bibnamefont {Smedback}},\ }\href {\doibase
  10.1088/1126-6708/2008/09/072} {\bibfield  {journal} {\bibinfo  {journal}
  {JHEP}\ }\textbf {\bibinfo {volume} {09}},\ \bibinfo {pages} {072} (\bibinfo
  {year} {2008})},\ \Eprint {http://arxiv.org/abs/0806.1519} {arXiv:0806.1519
  [hep-th]} \BibitemShut {NoStop}%
\bibitem [{\citenamefont {Halmagyi}\ \emph {et~al.}(2013)\citenamefont
  {Halmagyi}, \citenamefont {Petrini},\ and\ \citenamefont
  {Zaffaroni}}]{Halmagyi:2013sla}%
  \BibitemOpen
  \bibfield  {author} {\bibinfo {author} {\bibfnamefont {N.}~\bibnamefont
  {Halmagyi}}, \bibinfo {author} {\bibfnamefont {M.}~\bibnamefont {Petrini}}, \
  and\ \bibinfo {author} {\bibfnamefont {A.}~\bibnamefont {Zaffaroni}},\ }\href
  {\doibase 10.1007/JHEP08(2013)124} {\bibfield  {journal} {\bibinfo  {journal}
  {JHEP}\ }\textbf {\bibinfo {volume} {08}},\ \bibinfo {pages} {124} (\bibinfo
  {year} {2013})},\ \Eprint {http://arxiv.org/abs/1305.0730} {arXiv:1305.0730
  [hep-th]} \BibitemShut {NoStop}%
\bibitem [{\citenamefont {Hristov}\ \emph {et~al.}(2011)\citenamefont
  {Hristov}, \citenamefont {Toldo},\ and\ \citenamefont
  {Vandoren}}]{Hristov:2011ye}%
  \BibitemOpen
  \bibfield  {author} {\bibinfo {author} {\bibfnamefont {K.}~\bibnamefont
  {Hristov}}, \bibinfo {author} {\bibfnamefont {C.}~\bibnamefont {Toldo}}, \
  and\ \bibinfo {author} {\bibfnamefont {S.}~\bibnamefont {Vandoren}},\ }\href
  {\doibase 10.1007/JHEP12(2011)014} {\bibfield  {journal} {\bibinfo  {journal}
  {JHEP}\ }\textbf {\bibinfo {volume} {12}},\ \bibinfo {pages} {014} (\bibinfo
  {year} {2011})},\ \Eprint {http://arxiv.org/abs/1110.2688} {arXiv:1110.2688
  [hep-th]} \BibitemShut {NoStop}%
\bibitem [{\citenamefont {Dall'Agata}\ and\ \citenamefont
  {Gnecchi}(2011)}]{DallAgata:2010ejj}%
  \BibitemOpen
  \bibfield  {author} {\bibinfo {author} {\bibfnamefont {G.}~\bibnamefont
  {Dall'Agata}}\ and\ \bibinfo {author} {\bibfnamefont {A.}~\bibnamefont
  {Gnecchi}},\ }\href {\doibase 10.1007/JHEP03(2011)037} {\bibfield  {journal}
  {\bibinfo  {journal} {JHEP}\ }\textbf {\bibinfo {volume} {03}},\ \bibinfo
  {pages} {037} (\bibinfo {year} {2011})},\ \Eprint
  {http://arxiv.org/abs/1012.3756} {arXiv:1012.3756 [hep-th]} \BibitemShut
  {NoStop}%
\bibitem [{\citenamefont {Cassani}\ \emph {et~al.}(2012)\citenamefont
  {Cassani}, \citenamefont {Koerber},\ and\ \citenamefont
  {Varela}}]{Cassani:2012pj}%
  \BibitemOpen
  \bibfield  {author} {\bibinfo {author} {\bibfnamefont {D.}~\bibnamefont
  {Cassani}}, \bibinfo {author} {\bibfnamefont {P.}~\bibnamefont {Koerber}}, \
  and\ \bibinfo {author} {\bibfnamefont {O.}~\bibnamefont {Varela}},\ }\href
  {\doibase 10.1007/JHEP11(2012)173} {\bibfield  {journal} {\bibinfo  {journal}
  {JHEP}\ }\textbf {\bibinfo {volume} {1211}},\ \bibinfo {pages} {173}
  (\bibinfo {year} {2012})},\ \Eprint {http://arxiv.org/abs/1208.1262}
  {arXiv:1208.1262 [hep-th]} \BibitemShut {NoStop}%
\bibitem [{\citenamefont {Maldacena}\ and\ \citenamefont
  {Nunez}(2001)}]{Maldacena:2000mw}%
  \BibitemOpen
  \bibfield  {author} {\bibinfo {author} {\bibfnamefont {J.~M.}\ \bibnamefont
  {Maldacena}}\ and\ \bibinfo {author} {\bibfnamefont {C.}~\bibnamefont
  {Nunez}},\ }\bibfield  {booktitle} {\emph {\bibinfo {booktitle}
  {{Superstrings. Proceedings, International Conference, Strings 2000, Ann
  Arbor, USA, July 10-15, 2000}}},\ }\href {\doibase 10.1142/S0217751X01003935,
  10.1142/S0217751X01003937} {\bibfield  {journal} {\bibinfo  {journal} {Int.
  J. Mod. Phys.}\ }\textbf {\bibinfo {volume} {A16}},\ \bibinfo {pages} {822}
  (\bibinfo {year} {2001})},\ \bibinfo {note} {[,182(2000)]},\ \Eprint
  {http://arxiv.org/abs/hep-th/0007018} {arXiv:hep-th/0007018 [hep-th]}
  \BibitemShut {NoStop}%
\bibitem [{\citenamefont {Gauntlett}\ \emph {et~al.}(2002)\citenamefont
  {Gauntlett}, \citenamefont {Kim}, \citenamefont {Pakis},\ and\ \citenamefont
  {Waldram}}]{Gauntlett:2001qs}%
  \BibitemOpen
  \bibfield  {author} {\bibinfo {author} {\bibfnamefont {J.~P.}\ \bibnamefont
  {Gauntlett}}, \bibinfo {author} {\bibfnamefont {N.}~\bibnamefont {Kim}},
  \bibinfo {author} {\bibfnamefont {S.}~\bibnamefont {Pakis}}, \ and\ \bibinfo
  {author} {\bibfnamefont {D.}~\bibnamefont {Waldram}},\ }\href {\doibase
  10.1103/PhysRevD.65.026003} {\bibfield  {journal} {\bibinfo  {journal} {Phys.
  Rev.}\ }\textbf {\bibinfo {volume} {D65}},\ \bibinfo {pages} {026003}
  (\bibinfo {year} {2002})},\ \Eprint {http://arxiv.org/abs/hep-th/0105250}
  {arXiv:hep-th/0105250 [hep-th]} \BibitemShut {NoStop}%
\bibitem [{\citenamefont {Hohm}\ and\ \citenamefont
  {Samtleben}(2014)}]{Hohm:2013uia}%
  \BibitemOpen
  \bibfield  {author} {\bibinfo {author} {\bibfnamefont {O.}~\bibnamefont
  {Hohm}}\ and\ \bibinfo {author} {\bibfnamefont {H.}~\bibnamefont
  {Samtleben}},\ }\href {\doibase 10.1103/PhysRevD.89.066017} {\bibfield
  {journal} {\bibinfo  {journal} {Phys.Rev.}\ }\textbf {\bibinfo {volume}
  {D89}},\ \bibinfo {pages} {066017} (\bibinfo {year} {2014})},\ \Eprint
  {http://arxiv.org/abs/1312.4542} {arXiv:1312.4542 [hep-th]} \BibitemShut
  {NoStop}%
\bibitem [{\citenamefont {Ciceri}\ \emph {et~al.}(2016)\citenamefont {Ciceri},
  \citenamefont {Guarino},\ and\ \citenamefont {Inverso}}]{Ciceri:2016dmd}%
  \BibitemOpen
  \bibfield  {author} {\bibinfo {author} {\bibfnamefont {F.}~\bibnamefont
  {Ciceri}}, \bibinfo {author} {\bibfnamefont {A.}~\bibnamefont {Guarino}}, \
  and\ \bibinfo {author} {\bibfnamefont {G.}~\bibnamefont {Inverso}},\ }\href
  {\doibase 10.1007/JHEP08(2016)154} {\bibfield  {journal} {\bibinfo  {journal}
  {JHEP}\ }\textbf {\bibinfo {volume} {08}},\ \bibinfo {pages} {154} (\bibinfo
  {year} {2016})},\ \Eprint {http://arxiv.org/abs/1604.08602} {arXiv:1604.08602
  [hep-th]} \BibitemShut {NoStop}%
\bibitem [{\citenamefont {Varela}(2015)}]{Varela:2015ywx}%
  \BibitemOpen
  \bibfield  {author} {\bibinfo {author} {\bibfnamefont {O.}~\bibnamefont
  {Varela}},\ }\href@noop {} {\  (\bibinfo {year} {2015})},\ \Eprint
  {http://arxiv.org/abs/1512.04943} {arXiv:1512.04943 [hep-th]} \BibitemShut
  {NoStop}%
\bibitem [{\citenamefont {Hohm}\ and\ \citenamefont
  {Samtleben}(2015)}]{Hohm:2014qga}%
  \BibitemOpen
  \bibfield  {author} {\bibinfo {author} {\bibfnamefont {O.}~\bibnamefont
  {Hohm}}\ and\ \bibinfo {author} {\bibfnamefont {H.}~\bibnamefont
  {Samtleben}},\ }\href {\doibase 10.1007/JHEP01(2015)131} {\bibfield
  {journal} {\bibinfo  {journal} {JHEP}\ }\textbf {\bibinfo {volume} {1501}},\
  \bibinfo {pages} {131} (\bibinfo {year} {2015})},\ \Eprint
  {http://arxiv.org/abs/1410.8145} {arXiv:1410.8145 [hep-th]} \BibitemShut
  {NoStop}%
\bibitem [{\citenamefont {Dall'Agata}\ \emph {et~al.}(2012)\citenamefont
  {Dall'Agata}, \citenamefont {Inverso},\ and\ \citenamefont
  {Trigiante}}]{Dall'Agata:2012bb}%
  \BibitemOpen
  \bibfield  {author} {\bibinfo {author} {\bibfnamefont {G.}~\bibnamefont
  {Dall'Agata}}, \bibinfo {author} {\bibfnamefont {G.}~\bibnamefont {Inverso}},
  \ and\ \bibinfo {author} {\bibfnamefont {M.}~\bibnamefont {Trigiante}},\
  }\href {\doibase 10.1103/PhysRevLett.109.201301} {\bibfield  {journal}
  {\bibinfo  {journal} {Phys.Rev.Lett.}\ }\textbf {\bibinfo {volume} {109}},\
  \bibinfo {pages} {201301} (\bibinfo {year} {2012})},\ \Eprint
  {http://arxiv.org/abs/1209.0760} {arXiv:1209.0760 [hep-th]} \BibitemShut
  {NoStop}%
\bibitem [{\citenamefont {Inverso}(2016)}]{Inverso:2015viq}%
  \BibitemOpen
  \bibfield  {author} {\bibinfo {author} {\bibfnamefont {G.}~\bibnamefont
  {Inverso}},\ }\href {\doibase 10.1007/JHEP03(2016)138} {\bibfield  {journal}
  {\bibinfo  {journal} {JHEP}\ }\textbf {\bibinfo {volume} {03}},\ \bibinfo
  {pages} {138} (\bibinfo {year} {2016})},\ \Eprint
  {http://arxiv.org/abs/1512.04500} {arXiv:1512.04500 [hep-th]} \BibitemShut
  {NoStop}%
\bibitem [{\citenamefont {Borghese}\ \emph {et~al.}(2013)\citenamefont
  {Borghese}, \citenamefont {Dibitetto}, \citenamefont {Guarino}, \citenamefont
  {Roest},\ and\ \citenamefont {Varela}}]{Borghese:2012zs}%
  \BibitemOpen
  \bibfield  {author} {\bibinfo {author} {\bibfnamefont {A.}~\bibnamefont
  {Borghese}}, \bibinfo {author} {\bibfnamefont {G.}~\bibnamefont {Dibitetto}},
  \bibinfo {author} {\bibfnamefont {A.}~\bibnamefont {Guarino}}, \bibinfo
  {author} {\bibfnamefont {D.}~\bibnamefont {Roest}}, \ and\ \bibinfo {author}
  {\bibfnamefont {O.}~\bibnamefont {Varela}},\ }\href {\doibase
  10.1007/JHEP03(2013)082} {\bibfield  {journal} {\bibinfo  {journal} {JHEP}\
  }\textbf {\bibinfo {volume} {1303}},\ \bibinfo {pages} {082} (\bibinfo {year}
  {2013})},\ \Eprint {http://arxiv.org/abs/1211.5335} {arXiv:1211.5335
  [hep-th]} \BibitemShut {NoStop}%
\bibitem [{Gua()}]{Guarino:wip}%
  \BibitemOpen
  \href@noop {} {\ }\BibitemShut {NoStop}%
Work in progress
\bibitem [{\citenamefont {Lee}\ \emph {et~al.}(2015)\citenamefont {Lee},
  \citenamefont {Strickland-Constable},\ and\ \citenamefont
  {Waldram}}]{Lee:2015xga}%
  \BibitemOpen
  \bibfield  {author} {\bibinfo {author} {\bibfnamefont {K.}~\bibnamefont
  {Lee}}, \bibinfo {author} {\bibfnamefont {C.}~\bibnamefont
  {Strickland-Constable}}, \ and\ \bibinfo {author} {\bibfnamefont
  {D.}~\bibnamefont {Waldram}},\ }\href@noop {} {\  (\bibinfo {year} {2015})},\
  \Eprint {http://arxiv.org/abs/1506.03457} {arXiv:1506.03457 [hep-th]}
  \BibitemShut {NoStop}%
\bibitem [{\citenamefont {Cremmer}\ and\ \citenamefont
  {Julia}(1979)}]{Cremmer:1979up}%
  \BibitemOpen
  \bibfield  {author} {\bibinfo {author} {\bibfnamefont {E.}~\bibnamefont
  {Cremmer}}\ and\ \bibinfo {author} {\bibfnamefont {B.}~\bibnamefont
  {Julia}},\ }\href {\doibase 10.1016/0550-3213(79)90331-6} {\bibfield
  {journal} {\bibinfo  {journal} {Nucl. Phys.}\ }\textbf {\bibinfo {volume}
  {B159}},\ \bibinfo {pages} {141} (\bibinfo {year} {1979})}\BibitemShut
  {NoStop}%
\bibitem [{\citenamefont {de~Wit}\ \emph {et~al.}(2007)\citenamefont {de~Wit},
  \citenamefont {Samtleben},\ and\ \citenamefont {Trigiante}}]{deWit:2007mt}%
  \BibitemOpen
  \bibfield  {author} {\bibinfo {author} {\bibfnamefont {B.}~\bibnamefont
  {de~Wit}}, \bibinfo {author} {\bibfnamefont {H.}~\bibnamefont {Samtleben}}, \
  and\ \bibinfo {author} {\bibfnamefont {M.}~\bibnamefont {Trigiante}},\ }\href
  {\doibase 10.1088/1126-6708/2007/06/049} {\bibfield  {journal} {\bibinfo
  {journal} {JHEP}\ }\textbf {\bibinfo {volume} {0706}},\ \bibinfo {pages}
  {049} (\bibinfo {year} {2007})},\ \Eprint {http://arxiv.org/abs/0705.2101}
  {arXiv:0705.2101 [hep-th]} \BibitemShut {NoStop}%
\end{thebibliography}%

\end{document}